\documentclass[twocolumn]{aastex62}

\usepackage{graphicx}
\usepackage{dcolumn}
\usepackage{bm}
\usepackage{threeparttable}
\usepackage{enumitem} 
\usepackage{hyperref}

\received{\today}
\revised{xx, xx, 2018}

\shorttitle{X-ray morphology of Mars \thanks{ }}
\shortauthors{}

\begin{document}
\title{X-ray morphology due to charge-exchange emissions used to study the global structure around Mars}
\correspondingauthor{G.Y. Liang}
\email{gyliang@bao.ac.cn}

\author{G.Y. Liang}
\affiliation{CAS Key Laboratory of Optical Astronomy, National Astronomical Observatories, Chinese Academy of Sciences, Beijing 100101, China}
\affiliation{Institute for Frontiers in Astronomy and Astrophysics, Beijing Normal University, Beijing 102206, China}

\author{T.R. Sun}
\affiliation{National Space Science Center, Chinese Academy of Sciences, Beijing, China }

\author{H.Y. Lu}
\affiliation{School of Space and Environment, Beihang University, Beijing 100191, China}

\author{X.L. Zhu}
\affiliation{Institute of Modern Physics, Chinese Academy of Sciences, Lanzhou 730000, China}

\author{Y. Wu}
\affiliation{Key Laboratory of Computational Physics, Institute of Applied Physics and Computational Mathematics, Beijing 100088, China }

\author{S.B. Li$^4$} \noaffiliation

\author{H.G. Wei$^1$, D.W. Yuan$^1$} \noaffiliation

\author{J.Y. Zhong$^2$}
\affiliation{Department of Astronomy, Beijing Normal University, Beijing 100875, China } 

\author{W. Cui}
\affiliation{Department of Astronomy, Tsinghua University, Beijing 100084, China   }

\author{X.W. Ma$^5$, G. Zhao$^1$} \noaffiliation

\begin{abstract}
Soft x-ray emissions induced by  solar wind ions that collide with neutral material in the solar system have been detected around planets, and were proposed  as a remote probe for the solar wind interaction  with the Martian exosphere. A multi-fluid three-dimensional magneto-hydrodynamic model is adopted to derive the global distributions of solar wind particles. Spherically symmetric exospheric H, H$_2$, He, O, and CO$_2$ density profiles and a sophisticated hybrid model that includes charge-exchange and proton/neutral excitation processes are used to study the low triplet line ratio $G=\frac{i+f}{r}$ (0.77$\pm$0.58) of O VII and total x-ray luminosity around Mars. We further calculate the emission factor $\alpha$-value with different neutrals over a wide ion abundance and velocity ranges. Our results are in good agreement with those of previous reports. The evolution of the charge stage of solar wind ions shows that sequential recombination due to charge-exchange can be negligible at the interaction region. This only appears below the altitude of 400~km. The anonymous low disk $G$ ratio can be easily explained by the collisional quenching effect at neutral densities higher than 10$^{11}$cm$^{-3}$. However, the quenching contribution is small in Mars' exosphere and only appears below 400~km. Charge-exchange with H$_2$ and N$_2$ is still the most likely reason for this low $G$-ratio.  X-ray  emissivity maps in collisions with different neutrals differ from each other. A clear bow shock in the collision with all the neutrals is in accordance with previous reports. The resulting total x-ray luminosity of 6.55~MW shows a better agreement with the XMM-Newton observation of 12.8$\pm$1.4~MW than that of previous predictions.
\end{abstract}


\section{Introduction}
Solar wind ion--induced charge-exchange (SWCX) x-ray emissions have been detected from most planets in  the solar system, e.g., Earth~\citep{SCK04}), Mars~\citep{DLB06}, Saturn~\citep{BBE10}, Jupiter~\citep{BBE07,HSK09}. A review of SWCX emission in the solar system was presented by \cite{DLB12}.  This kind of x-ray emission was further suggested as a probe for remote monitoring of the magnetosheath and magnetopause of the planets and the solar wind by \cite{SCC09,SWS19} for Earth. \cite{GHK04} estimated the SWCX total x-ray luminosity (1.8~MW) to be consistent with the Chandra 2001 observation for Mars that used a hybrid model for the solar wind--Mars interaction and a test particle simulation of heavy ion trajectories near Mars, where a constant cross-section for H and O atoms and a simplified two-step cascade model were adopted. Later, \cite{KMC12}  performed a three-dimensional (3D) multi-species hybrid simulation model, and reproduced the solar wind--Martian interfaces well by using x-ray morphology, and confirmed its feasibility  for the remote probing to the interaction between the solar wind and the Martian exosphere. However, the estimated luminosity of $\sim$0.7--2.0~MW is far smaller than the total observed luminosity of 12.8$\pm$1.4~MW~\citep{DLB06}. \cite{KMC12}  included three neutral profiles of H, O, and CO$_2$ in the Martian environment and constant cross-sections with H and H$_2$O without the H$_2$ and He components. Additionally, an anonymous low line intensity ratio $(i+f)/r$ of O VII in disk observation cannot be explained by the charge-exchange process. 

In this work we incorporate recent progress on solar wind ions distributions from a multi-fluid magneto-hydrodynamic (MHD) simulation comparing with the Mars Atmosphere and Volatile Evolution (MAVEN) spacecraft data. Furthermore, we include more neutral components in the Martian exosphere to re-study the x-ray luminosity. In section 2 we describe the MHD model for solar wind ions, the neutral profile, as well as a sophisticated hybrid atomic model for x-ray emission. In section 3 we present the temperature- and velocity-dependent $\alpha$-value, the evolution of charge-stage of solar wind ions, the collisional quenching effect on line ratios of O VII, and the x-ray emissivity maps in the XZ plane. Finally, a summary and conclusion are outlined in section 4.

\section{Simulation model}
In this interaction of solar wind ions with Martian neutrals (including H, H$_2$, He, O, CO$_2$), the local volume emission rate $P(r)$ from $X^{q+}$ charged ion is given by {\cite{Cra97,KLK09},
\begin{eqnarray}
P^{q+}(r) = N^{q+}(r)\sum_{ij,k}\epsilon^k_{ij}(v),
\end{eqnarray}
where $N^{q+}$ is number density of solar wind ions $X^{q+}$ at given location ($r,\theta$, $\phi$), respectively. $\epsilon^k_{ij}(v)$ is the $i\to j$ line emissivity at a relative collisional velocity $v=\sqrt{v_{sw} + 3k_BT/m_p}$ between solar wind ion and $k$-th neutral. Here $v_{sw}$ is the bulk velocity of the solar wind and the second term is the general thermal velocity. The number density of different charged ions $X^{q+}$ is a self-consistent solution to the differential equation:
\begin{eqnarray}
\frac{dN^{q+}(r)}{dr} & = - N^{q+} \sum_k\sigma_k^{q+}(v)n^k_{\rm neu}  \\ \nonumber
                                 &  + N^{(q+1)+}\sum_k \sigma_k^{(q+1)+}(v)n^k_{\rm neu} 
\end{eqnarray}
that describes the evolution of solar wind charged ions in the radial direction, where the initial  $X^{q+}$ ion number density $N^{q+}(r_{\infty})=Ab*n_{\rm sw}$ is defined by its abundance in the solar wind before interaction.
 
\subsection{Neutral atmosphere and exosphere}
The Martian neutral environment is composed of CO$_2$, O, H, H$_2$, He, and N$_2$ as shown in Fig.~\ref{fig-mars-neutrals}.  The density profiles of CO$_2$, and O are fit to results from the Mars thermosphere global circulation model (MTGCM) as reported by \cite{KMC12}. 
The hydrogen density profile  is fit to results from \cite{AH71} and \cite{Kra02} and can be written by
\begin{eqnarray}
 \nonumber n_{\rm H} & = 10^3 {\rm exp}\left[9.25\times10^5\left( \frac{1}{z+3393.5} - \frac{1}{3593.5} \right )  \right] \\ 
        &  + 3.0\times10^4 {\rm exp}\left[1.48\times10^4 \left( \frac{1}{z+3393.5} - \frac{1}{3593.5} \right )  \right ]
\end{eqnarray}

For the H$_2$ and He density profile we adopt the results of \cite{Kra10}with a similar fitting formula as H but with different fitting parameters as given by
\begin{eqnarray}
 \nonumber n_{\rm H_2} & = 1.59\times10^4 {\rm exp}\left[1.48\times10^6\left( \frac{1}{z+3384.1} - \frac{1}{3548.2} \right )  \right] \\ 
        &  + 5\times10^4 {\rm exp}\left[4.8\times10^4 \left( \frac{1}{z+3384.1} - \frac{1}{3548.2} \right )  \right ], 
\end{eqnarray}

\begin{eqnarray}
 \nonumber n_{\rm He} & = 1.79\times10^4 {\rm exp}\left[1.38\times10^6\left( \frac{1}{z+3381.1} - \frac{1}{3549.2} \right )  \right] \\ 
        &  + 9\times10^4 {\rm exp}\left[8.5\times10^4 \left( \frac{1}{z+3381.1} - \frac{1}{3549.2} \right )  \right ].
\end{eqnarray}
The fitting values at low altitudes are consistent with the results of \cite{Kra10}.
 
 Recently, these neutral profiles below 500~km have been measured by the Neutral Gas and Ion Mass Spectrometer (NGIMS) on the Mars Atmosphere and Volatile EvolutioN (MAVEN) mission \citep{MBE15,SYB22,WCN21}. The hydrogen profile was measured via the 1216~\AA\, Lyman alpha line covered by the Imaging Ultraviolet Spectrograph (IUVS) on the MAVEN mission~\citep{CCD18}. \cite{SYB22} further investigated their horizontal variations with local time, latitude and season. The magnitude of variation is very large, and even more than an order of magnitude for H$_2$ and He \citep[see Fig.~6 there]{SYB22}. For comparison, the MAVEN measurements during the orbit period of DD2~\citep{WCN21,SYB22} and similar fitting as above are presented in the bottom panel of Fig.~\ref{fig-mars-neutrals}. Since there are not in-situ measurement data for the neutral density above 1000~km, we still can not calibrate the uncertainty of this extrapolation. Yet it approximately follows the relation of $\sim1/r^3n^{\dag}_{\rm H}$ (here $n^{\dag}_{\rm H}$ is the hydrogen density near Mars). Although their proportional relation to the x-ray emission as shown in Eq.(2), the low neutral density at distant halo region makes the resultant total X-ray luminosity to be affected smalle by the uncertainty of this extrapolation. By using the two different groups of the neutral profiles, we will estimate the uncertainty of the total luminosity from neutrals, that will be discussed in the following discussion section.

\begin{figure}
   \centering
   \includegraphics[width=9cm, angle=0]{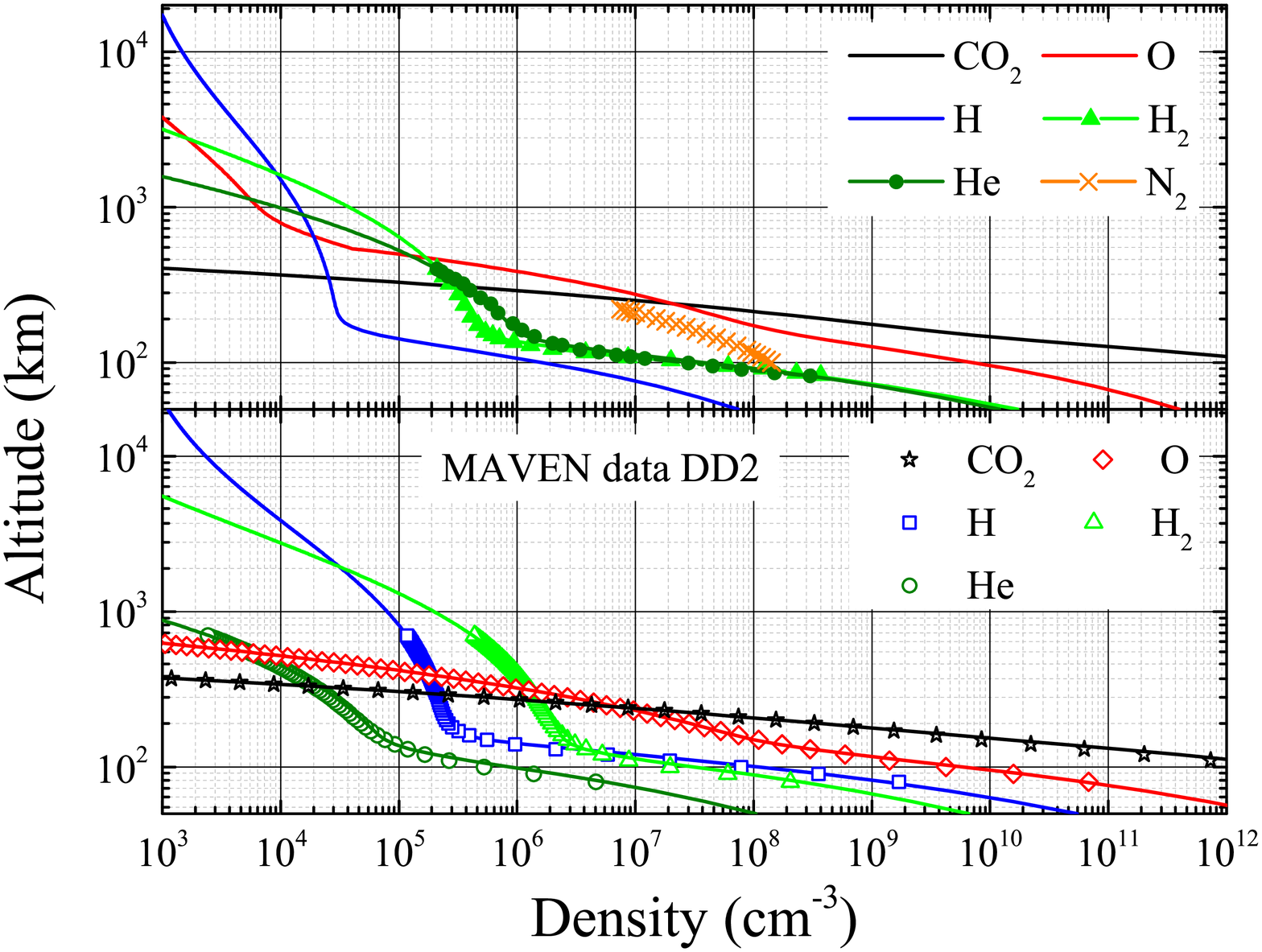} %
   \caption{Neutral density profile of different species in the Martian atmosphere. {\it Upper:} For H$_2$ and He neutrals, a fitting procedure was used to extend the available data (symbols) from \cite{Kra10} to a higher altitude. For N$_2$ gas, the density profile is from the work of \cite{BPB15}. {\it Bottom:}  The MAVEN measurement at the period of DD2 \citep{WCN21} and the corresponding fitting. } 
   \label{fig-mars-neutrals}
\end{figure}

\subsection{Global distribution of solar wind ions around Mars}
The interaction of the solar wind with the Martian atmosphere and ionosphere results in the formation of two distinct boundaries, i.e., bow shock and magnetic pileup (or magnetic pause). \cite{MCD05,MCD06} investigated the global structure around Mars  using a three-dimensional multi-species (proton and He$^{2+}$) hybrid model, where complicated substructures were demonstrated in density maps of protons and $\alpha$-particles. Using the MAVEN spacecraft data and an algorithm of automated region identification, \cite{NLN20}  derived an empirical model for both the bow shock and magnetic pileup boundary locations around Mars. By assuming cylindrical symmetry, a parabolic model with a focus on the subsolar standoff location ($x_0=C~p^{\beta}_{sw}F^{\gamma}B^{\delta}$) in the $x$-axis (Sun-Mars center line) can well fit the two boundaries, i.e.,
\begin{eqnarray}
\rho^2=& \alpha(x-x_0).
\label{eq-bowshock}
\end{eqnarray}
Here, $\rho$ and $x$ (in $R_{\rm M}$) are aberration-corrected cylindrical coordinates of a given point at the bow shock and magnetic pileup boundary. $C$, $\alpha$, $\beta$, $\delta$, and $\gamma$ are parameters fit to the MAVEN spacecraft data for the bow shock and magnetic pileup boundary.  $p_{sw}$ is dynamic pressure in nPa, $F$ refers to the solar ionizing flux in mW m$^{-2}$, and $B$ denotes the magnitude of the magnetic field in n$T$. The best-fit result for the bow shock and magnetic pileup boundary are illustrated by thick black curves in Fig.~\ref{fig-swden}. 


In this work a multi-fluid 3D MHD model is adopted to derive the global distributions of the solar wind particle (H$^+$) density and velocity surrounding Mars, where the Navier-Stoker equations for five significant ion species (i.e., proton in solar wind, H$^+$ from Mars, O$_2^+$, O$^+$, CO$_2^+$)  in the Martian ionosphere were used to describe the physics. The equations include conservation equations for the plasma flow with respect to continuity, momentum, and energy. The Navier-Stoker equations for each species are augmented by the interaction of the electromagnetic effects. The physical detail for this 3D MHD model can be found in the studies of \cite{LLC20,LLC22}.

The computational domain is -24$R_{\rm M} \leq x$ $\leq 8R_{\rm M}$, -16$R_{\rm M} \leq y/z$ $\leq 16R_{\rm M}$, where $R_{\rm M}$ is the radius of Mars ($R_{\rm M}=3396$ km). Because the general curvilinear coordinate system is adopted, a high resolution for the region with the most intense variations of physical parameters is achieved by refining the physical grid, and the smallest grid size can approach 60 km. The solar wind density and velocity are chosen to be 4 cm$^{-3}$ and 500 km/s, respectively. The interplanetary magnetic field (IMF) is chosen to be 3 nT. 

The solar wind density and velocity profiles in the XZ-plane show a clear bow shock with its position being excellent agreement  with that from MAVEN spacecraft data \citep{NLN20}, see Fig.~\ref{fig-swden}.   

\begin{figure*}[ht]
   \centering
     \includegraphics[width=5.5cm, angle=0]{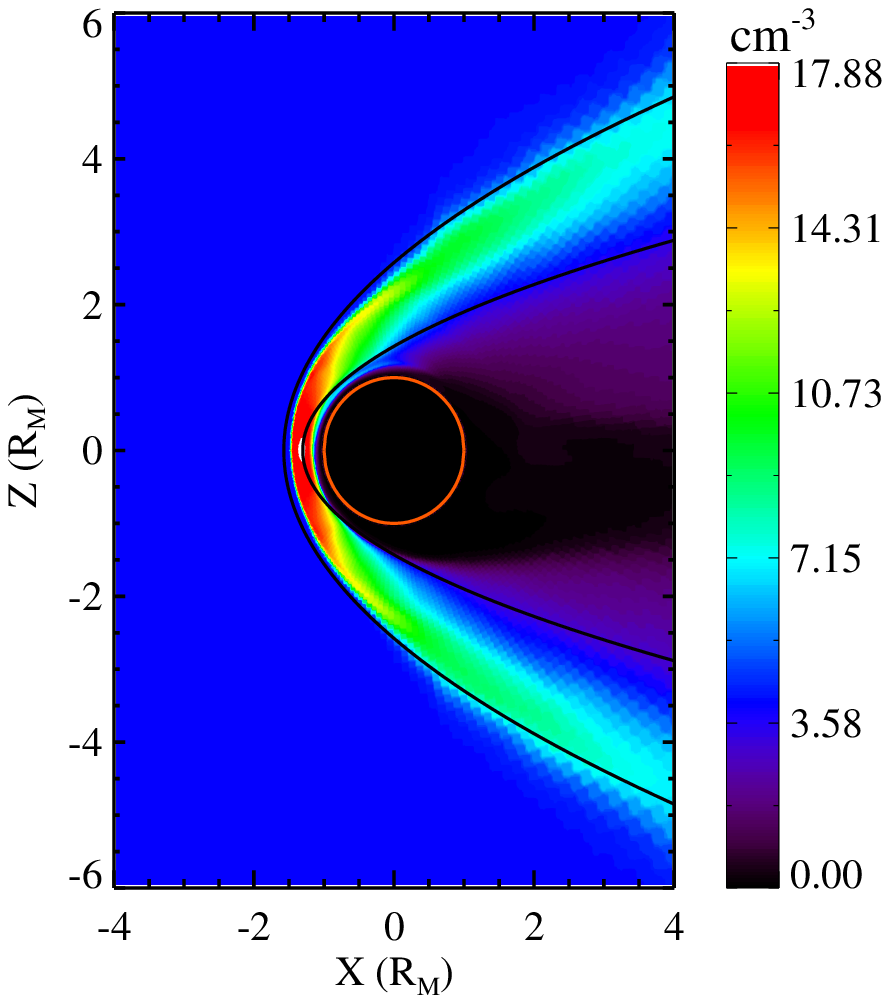}
      \includegraphics[width=5.5cm, angle=0]{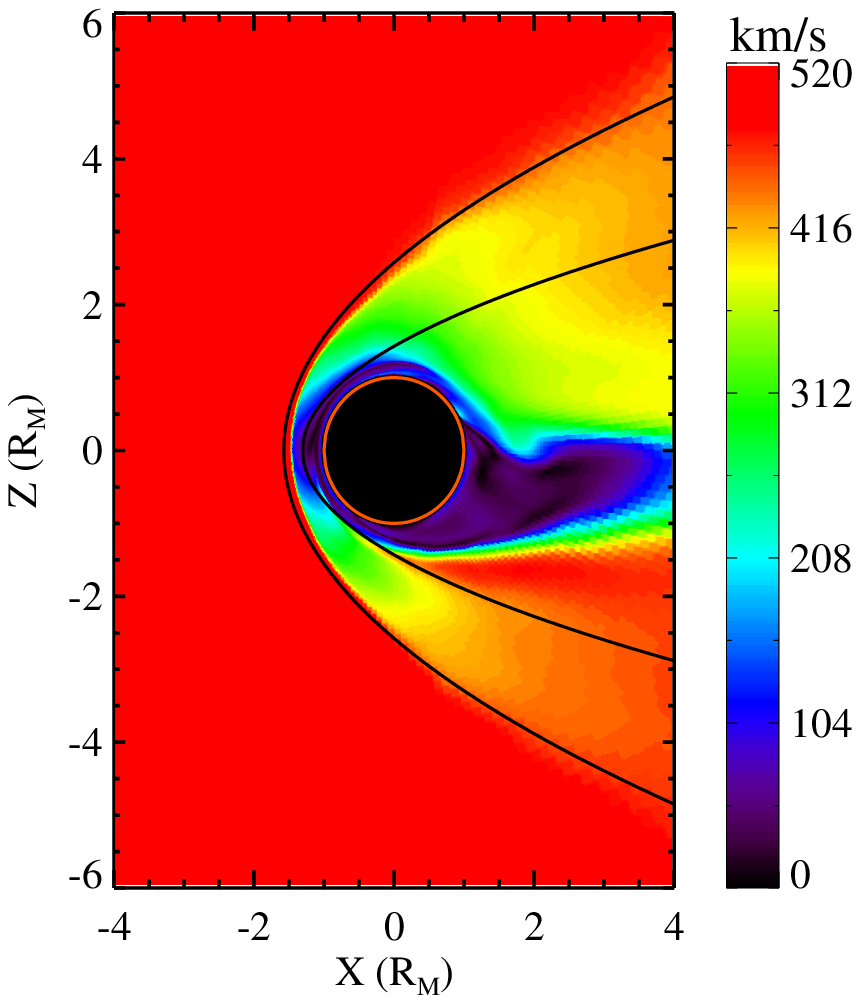}
   \caption{{\it Left:}  Density map of the solar wind H$^+$ species reproduced from a 3D five species multi-fluid magnetohydrodynamic (MHD) simulation by \cite{LLC20}. {\it Right:} Velocity map of the solar wind ions reproduced from the 3D multi-fluid MHD simulation by \cite{LLC20,LLC22}. The thick solid curves represent the locations of bow shock and magnetic pileup boundary from the formula derived by \cite{NLN20} based on the MAVEN spacecraft data. 
   } 
   \label{fig-swden}
\end{figure*}

\subsection{Emission model}
To calculate the line emissivity $\epsilon$ of $X^{q+}$ at a given location ($r, \theta$, $\phi$) due to charge-exchange electron captures with neutrals, and to calculate subsequent radiative decays, either directly to the ground and lower excited states or via cascades, we obtain the density $N^{q+}_i$ of an ion $X^{q+}$ at a given $i$-th level state by solving the following rate equation assuming equilibrium:
\begin{eqnarray}
\nonumber \frac{d}{dt}N_i^{q+}(r)  = \sum_{j>i}N_j^{q+}(r)A_{ji} - \sum_{j<i}N_i^{q+}(r)A_{ij} \\
 + \sum_kn^k_{\rm neu}(r)N_0^{(q+1)+}\left[ C^k_{0i}(v) + \eta^{2\to1}C^{2,k}_{0i}(v)   \right] \\
\nonumber + n_{\rm H}\sum_{j\neq i}\left[ N_j^{q+}(r)Q_{ji}(T_{\rm e}) - N_i^{q+}(r)Q_{ij}(T_{\rm e}) \right] \\
\nonumber  = 0, 
\label{eq-pop}
\end{eqnarray}
where $N_i^{q+}$ is the number density of $q+$ charged ions at $i$-th level state, while $n^k_{\rm neu}$ corresponds to the number density of $k$ kind of neutrals, e.g. H, H$_2$, He, O, and CO$_2$. $n_{\rm H}$ refers to the density of H.  $C^k_{ij}\equiv<v\sigma^k_{\rm cx}(v)>$ and $C^{2,k}_{ij}\equiv<v\sigma^{2,k}_{\rm cx}(v)>$ are the single and double-electron capture rate coefficients, respectively. $A_{ij}$ is the radiative decay rate for a given transition line $i\to j$. $Q_{ij}(T_{\rm e})$ are the proton impact excitation rates at a given temperature $T_{\rm e}$ that are used to account for neutral impact excitations due to high density at lower altitudes, and are only considered for O~VII triplets at disk observation.  $v$ is the relative collision velocity between $X^{(q+1)+}$ and $k$ kind of neutrals, while $\sigma^k_{\rm cx}(v)$ and $\sigma^{2,k}_{\rm cx}(v)$ are the cross-section of single- and double-electron transfer processes. $\eta^{2\to1}$ refers to the ratio of ionic fraction between $(q+2)+$ and $(q+1)+$ charged ions before electron capture. Furthermore, the line emissivity $\epsilon_{ij}=N^{q+}_iA_{ij}$ can be obtained.

The atomic data of level energies and radiative decays have been reported by \cite{LLW14,LZW21}. Here the  charge-exchange cross-sections are from the Kronos v3.1 database~\footnote{www.physast.uga.edu/research/stancil-group/atomic-molecular-databases/kronos \label{ft_kronos}} that is implemented by Stancil research group in a series of studies~\citep{CHS14,CML18,MCL16,MCL17} that use multiple methods, including multichannel Landau-Zener (MCLZ), atomic-orbital close-coupling (AOCC), molecular-orbital close-coupling (MOCC), and classical trajectory Monte Carlo methods. For the collision of O$^{7+}$ with the O atom, there are no data available. Hence, the MCLZ CX cross-section of O$^{7+}$ with water (H$_2$O) is used because of their similar weight. For the double-electron capture cross-section, we adopt those available data that are explained in the study by \cite{LZW21}. Since the data availability from AOCC/MOCC method are in the Kronos database, we adopt cross-sections from the AOCC/MOCC method for  bare- and H-like ion collisions with the hydrogen or helium atom, while data from the MCLZ method will be used for collisions with other neutrals. This means the velocity-dependent CX cross-section is used in this work, not the constant cross-section used in the work of \cite{KMC12}. For example, Fig.~\ref{fig-cs-o8-neus} shows the cross-section of H-like oxygen O$^{7+}$ ion collisions with different neutrals in the Martian exosphere.  For comparison the values used by \cite{KMC12} are also  plotted; these values have a smaller cross-section with H and a larger cross-section with O than the accurate calculation in Kronos v3 database by $\sim$40\% and $\sim$300\% respectively, and were used by \cite{KMC12}. Therefore, the x-ray emissions would be under-/over-estimated in this previous work. In the collision with H, \cite{ZSA22} measured the absolute cross-sections at collisional velocities covering the typical solar wind velocities; see symbols with error bars in Fig.~\ref{fig-cs-o8-neus}. Thus, these experimental data in the collision with H are used in this work. There is an obvious difference between the experimental results and the theoretical calculations including those from Kronos database and from Gu's fitting below 200~km/s. This illustrates that a sophisticated method and laboratory measurements are still required for the velocity-dependent cross-sections in the collisions with other neutrals, even the best available cross-sections are used in this paper. This kind of uncertainty has a significant effect on spectral analyses for observations with high-resolution by using the charge-exchange model, and has been pointed out by \cite{GSZ22} for advanced theoretical calculations for especially the
low collision energy regime, in combination with more laboratory measurements.
\begin{figure}
   \centering
   \includegraphics[width=8.5cm, angle=0]{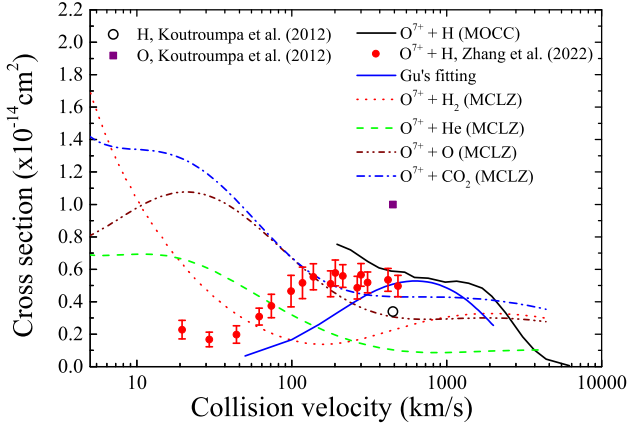} 
   \caption{Charge-exchange cross-section of O$^{7+}$ with H, H$_2$, He, O and CO$_2$,from the Kronos v3.1 database~$^{\ref{ft_kronos}}$. For the collision with the H atom, the data are from MOCC calculation, and Gu's fitting \citep{GKR16}. Symbols refer to the values used by \cite{KMC12} for collisions with H (open circle) and O (filled square). Symbols with error bars are from recent experiments by \cite{ZSA22}.
   } 
   \label{fig-cs-o8-neus}
\end{figure}

\section{Results and Discussion}
\subsection{Velocity and temperature dependence of $\alpha$ value}
The $\alpha$-value combined with MHD modeling is extensively used by the space physics community. This is a proportional factor based on a combination of the relative abundances and the cross-section of each possible interaction between a solar wind ion and a neutral particle $k$ causing line emissions as defined by \cite{Cra97} and \cite{WSC16} as follows:
\begin{eqnarray}
\alpha_k(X)=\sum_{q+}\alpha_k(X^{q+}) = \sum_{q+} \left[\frac{X^{q+}}{O}\right]\left[\frac{O}{H}\right] \epsilon_k(X^{q+}),
\end{eqnarray}
where ion emissivity {\bf $\epsilon_k(X^{q+})$} of a $q+$ charged ion is obtained by summing the line emissivity {\bf $\sum_{ij}\epsilon_{k,ij}(X^{q+})$} with the neutral particle of $n^k_{\rm neu}=1.0$ for a given transition $i\to j$. The line emissivity of one solar wind species {\bf $\epsilon_{k,ij}(X^{q+})\equiv N^{q+}_jA_{ij}$ }can be obtained from Eq.~(7) at the collision with the $k$ neutral  by multiplying by the radiative rate $A_{ij}$. In previous works \citep{SC00,Cra97,WS16}, the line emissivity {\bf $\sum_{ij}\epsilon_{k,ij}(X^{q+})$} in Eq.~(8) is replaced by $\sigma_{ij}\Delta E_{ij}$, which  means cascading effects to the upper atomic states $j$ of the $\Delta E_{ij}$ transition have been neglected. 

Eq.~(8) shows that the $\alpha$-value is highly variable depending on the ionic fraction of a given element and its abundance. Due to the extreme low electron density in inter-planetary space, the charge state distribution of solar wind ions will freeze-in after  leaving the solar surface \citep{LGL12}. This distribution in the solar wind provides insight into its origin with the characteristic temperature from the Sun.  Thus, we adopt the temperature in collisional equilibrium to represent the relative ionic fraction in solar wind. Figure~\ref{fig-alpha-o} shows the $\alpha$-values (in eV cm$^2$) of oxygen in collisions with different neutrals (e.g., H, H$_2$, He, O and CO$_2$) as a function of solar wind velocity and  logarithmic temperature (in the unit of K). Here, the solar abundance of \cite{LPG09} is adopted with oxygen to hydrogen abundance ratio of [O/H]=6.05$\times10^{-4}$, which is slightly higher than the mean values ranging from 2.03$\times10^{-4}$ to 4.76$\times10^{-4}$ from the ACE SWICS  data for fast and slow solar winds~\citep{WS16}, but consistent with the reported mean [O/H] ratio of 3.94$\times10^{-4}$ with a standard deviation of 3.01$\times10^{-4}$ by \cite{WSC16} from the OMNI data~\footnote{https://omniweb.gsfc.nasa.gov/}. By using the ACE (Advanced Composition Expoloer) data (including the ion density of O$^{7+, 8+}$, oxygen abundance [O/H], and solar wind velocity) obtained over 13 years (1998--2011) and the charge-exchange cross-sections from \cite{Bod07}, the resulting $\alpha$-value has a modal peak at $6\times10^{-16}$~eV cm$^2$ when colliding with hydrogen \citep[see Fig. 2]{WS16}. To compare with empirical methods \cite{WSC16} derived an $\alpha$-value of 7.6$\times10^{-16}$~eV~cm$^2$ by using [O/H]=1.1$\times10^{-3}$, an O$^{7+}$ abundance of 0.28, and an O$^{8+}$ abundance of 0.05. The present calculation (7.3$\times10^{-16}$~eV cm$^2$) shows an excellent agreement with the value measured at the temperature corresponding to the O$^{7+, 8+}$ abundance ratio at equilibrium. The present calculation of log($T_e$)/K=6.1 shows a good agreement with that of \cite{WS16} at a typical solar wind velocity of 300--600 km/s, where in-situ ACE data including velocity, O$^{7+,8+}$ abundance, and abundance ratio [O/H], were used. Both the present calculations and previous work from ACE data demonstrate that there is a strong dependence on the ionic fraction or temperature between log($T_e$)/K=6.0 and 6.5.
\begin{figure}[ht]
   \centering
   \includegraphics[width=8.5cm, angle=0]{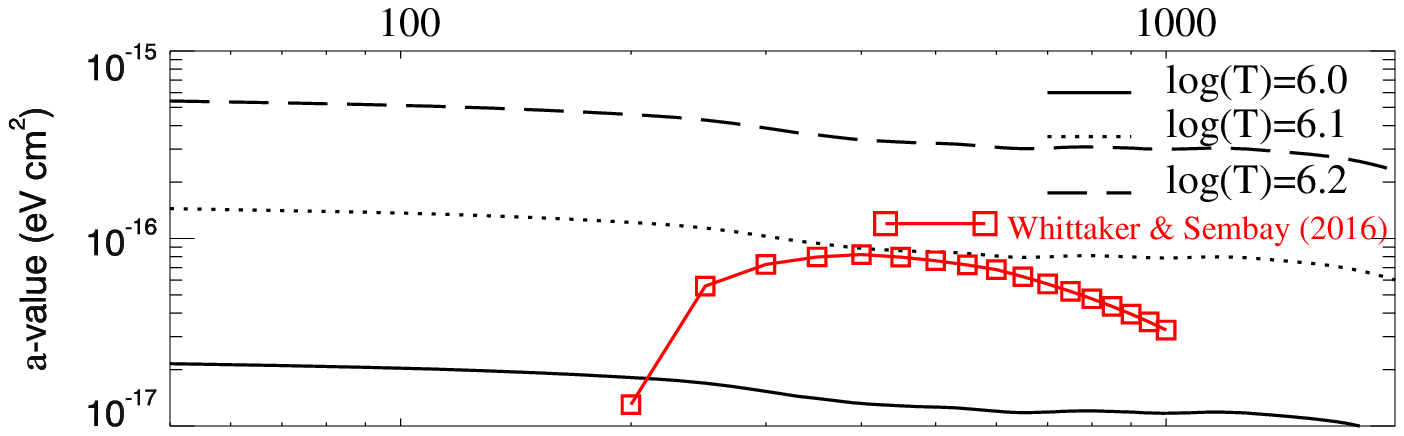}
   \includegraphics[width=8.5cm, angle=0]{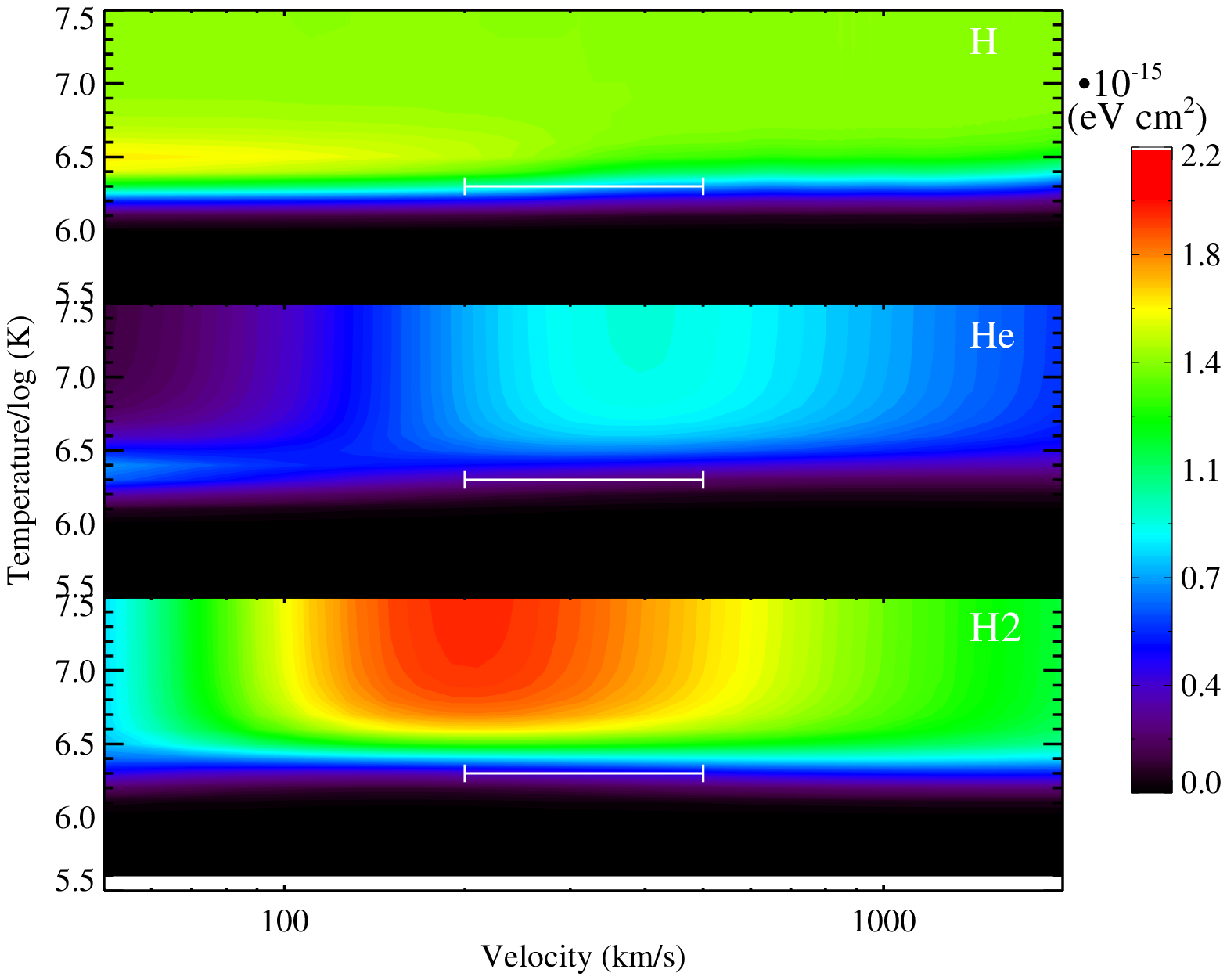}
   \includegraphics[width=8.5cm, angle=0]{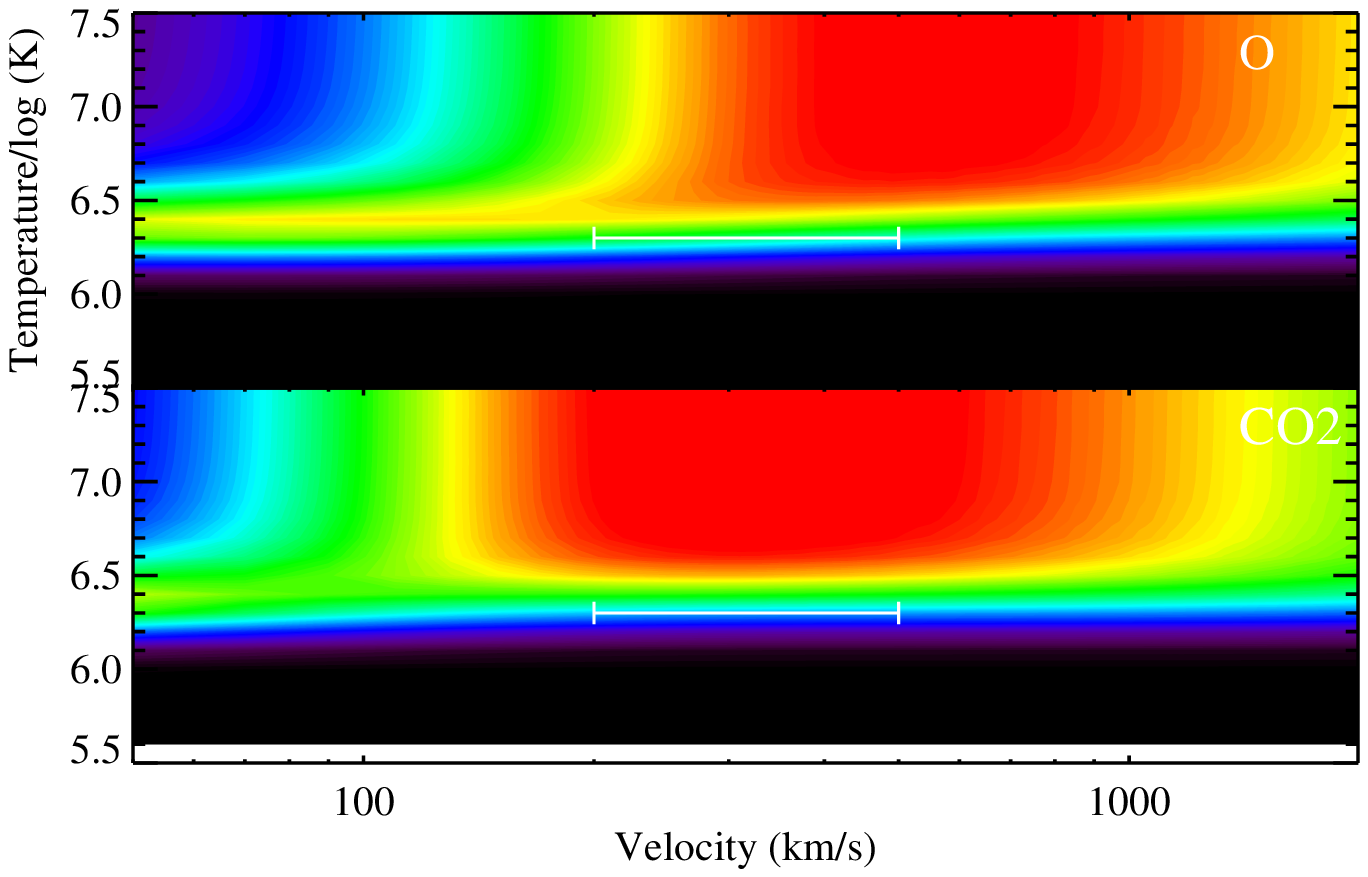}
   \caption{{\it Top:} Present $\alpha$-value of oxygen in collision with hydrogen at three temperatures (log($T_e$)/K=6.0, 6,1 and 6.2), along with calculated ones from the work of \cite{WS16})  based on parameters (velocity, O$^{7+,8+}$ fraction and abundance ratio [O/H]) from ACE data.  Contour plots for $\alpha$-values (eV cm$^2$) of oxygen in collisions with H, H$_2$, He, O and CO$_2$ in the temperature range log($T_{\rm}$) = 5.0--7.0 and velocity range of 50--2000 km/s with grid of 0.1 and 20 km/s, respectively. Here the  solar abundance for oxygen [O/H]=6.05$\times10^{-4}$ is used. The horizontal solid lines refer to the  typical solar wind velocity range of 200--500 km/s and the temperature (log($T)=6.3$ K)  with O$^{7+}$ and O$^{8+}$ abundances of 0.28 and 0.05 in collisional ionization equilibrium, respectively. } 
   \label{fig-alpha-o}
\end{figure}

 Hydrogen gas (H$_2$), He, O, and CO$_2$ are also important components in the Martian neutral environment. Both the multi-fluid 3D MHD computation performed by \cite{LLC20} and MAVEN data demonstrate that the bow shock and magnetic pileup boundary are approximately 1.55~$R_{M}$ (corresponding to the altitude of 1868 km \footnote{In the unit of the Martian radius $R_{\rm M}$, this value is relative to Mars' center, otherwise it refers to the altitude relative to the Martian surface.}) and 1.3~$R_{M}$ (1018 km) in the direction of Sun to Mars, respectively, see Fig.~\ref{fig-swden}. Below this altitude the number densities of H$_2$ and He become dominant and are higher than those of atomic hydrogen, see Fig.~\ref{fig-mars-neutrals}. We also present the $\alpha$-value in collision with other Martian neutrals. An  obvious dependence on solar wind velocity appears below $\sim$200-300 km/s. Large decrease of solar wind velocity after bow shock as shown in Fig~\ref{fig-swden} reveals that the constant $\alpha$-value adopted in previous works should generate large uncertainties \citep{WS16,WSC16}. The present $\alpha$ calculation will help improve the estimation of x-ray emission in the interaction region. In the calculated temperature and velocity grids the $\alpha$-value with H$_2$, O, and CO$_2$ is systematically higher than the value in the collision with H. 

We further calculate the $\alpha$-value of carbon, nitrogen and neon in the collisions with H, H$_2$, He, O, and CO$_2$, see Fig.~\ref{fig-alpha-cn}. Compared to the $\alpha$-value of oxygen, the carbon $\alpha$-value is lower than that of oxygen, while the nitrogen $\alpha$-value is even lower. This is mainly resultant from the higher abundance of oxygen than carbon and nitrogen in the solar wind. According to the mean abundance from ACE data of highly charged oxygen ions (O$^{8+}$/0.28, O$^{7+}$/0.05, \cite{WSC16}) and carbon ions (C$^{6+}$/0.13, C$^{5+}$/0.37, \cite{KMC12}), the calculated $\alpha$-values in the collisions with H are about 7.9$\times10^{-16}$~eV cm$^2$ (O) and 4.4$\times10^{-16}$~eV cm$^2$ (C), respectively. With the similar temperature of log$(T)=6.1\pm0.1$, the nitrogen $\alpha$-value is $\sim3.5\times10^{-16}$~eV cm$^2$. We also notice that there is a sudden decrease around 900~km/s for the $\alpha$-value of N in the collision with H. This is due to the recommended $nl$-selective cross-section used \citep[see Fig.~5 there]{WSL11}, and its complicated $nl$-distribution along the collisional velocity.

\begin{figure*}[h]
   \centering
   \includegraphics[width=8.5cm, angle=0]{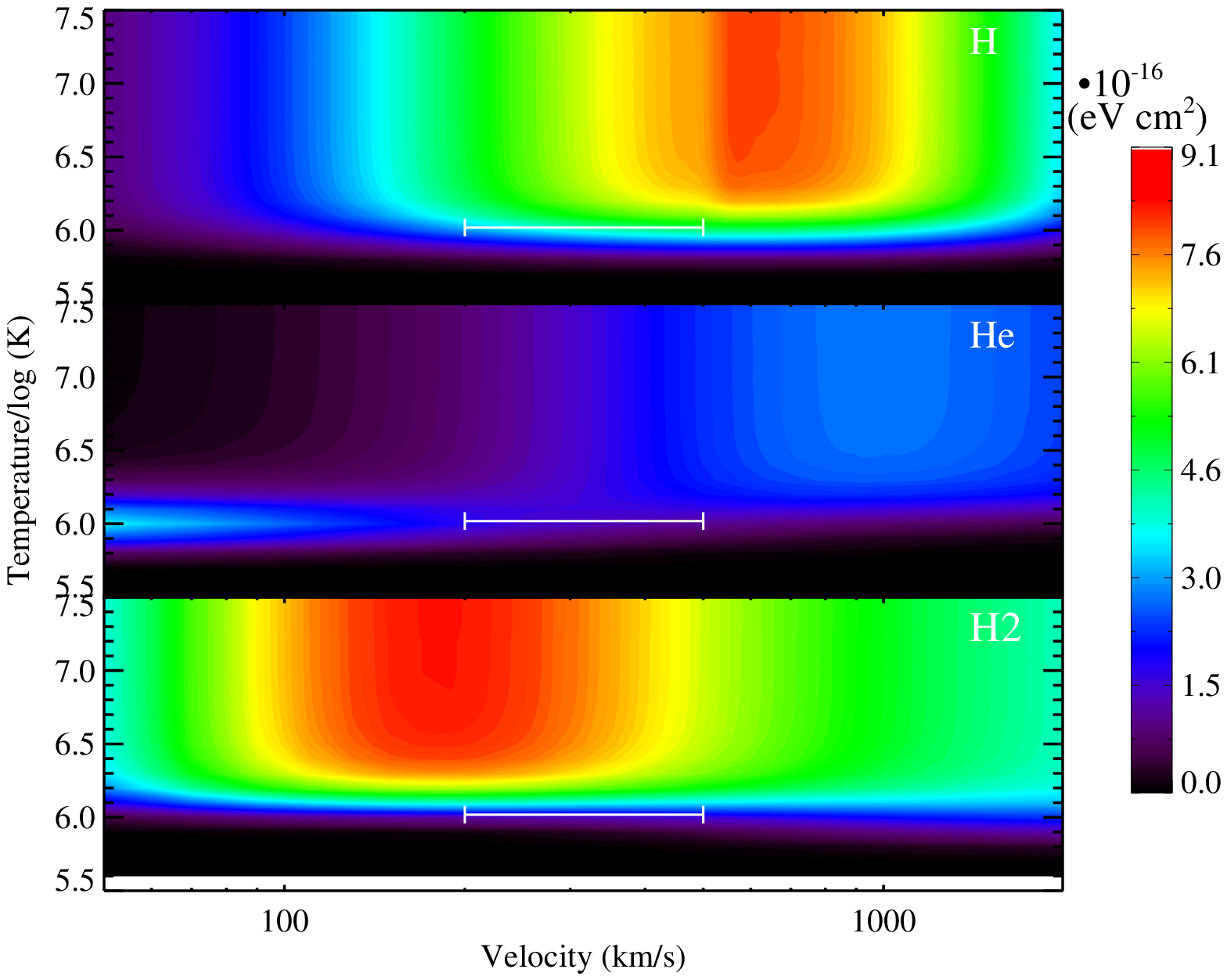}  \includegraphics[width=8.5cm, angle=0]{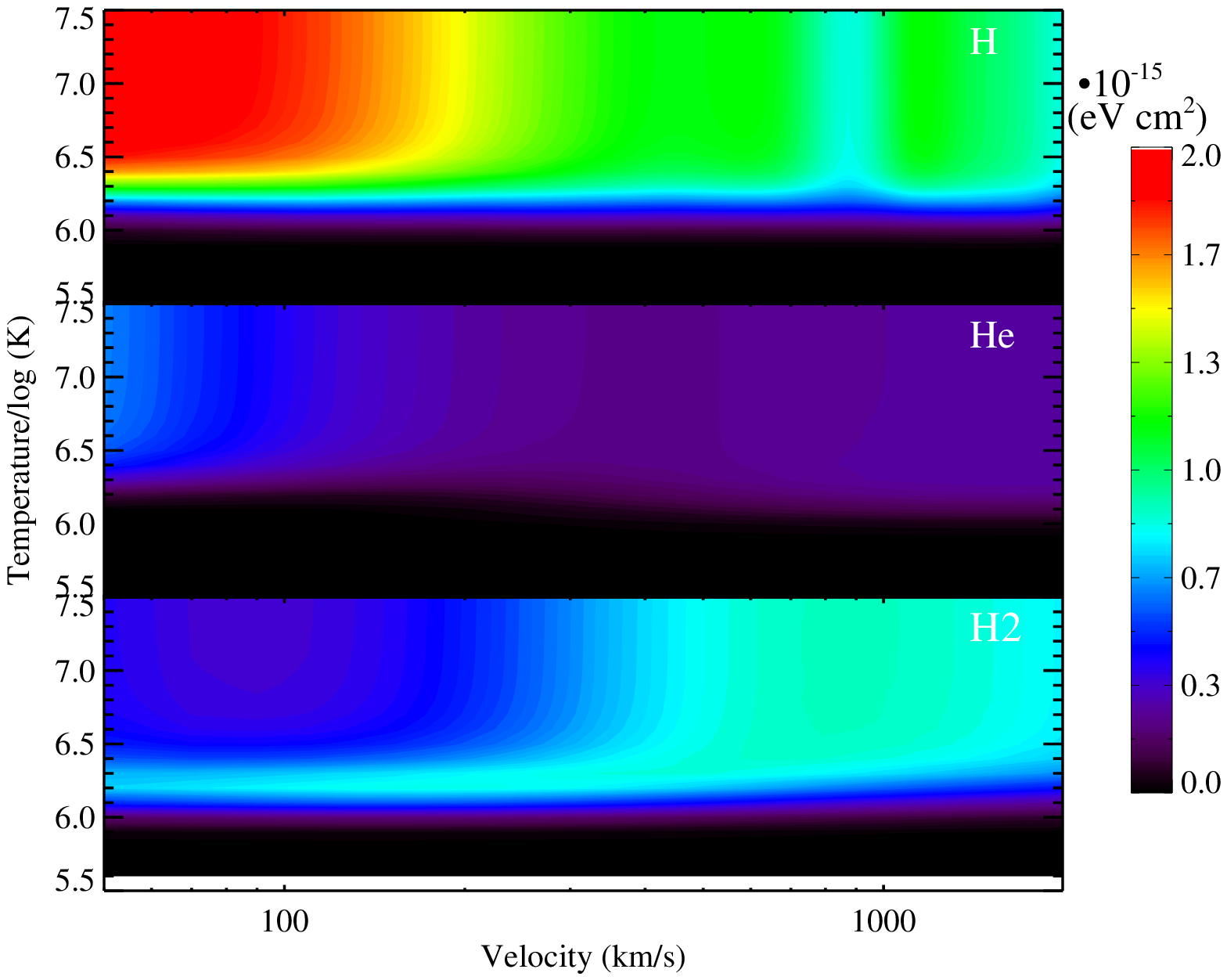} 
   \includegraphics[width=8.5cm, angle=0]{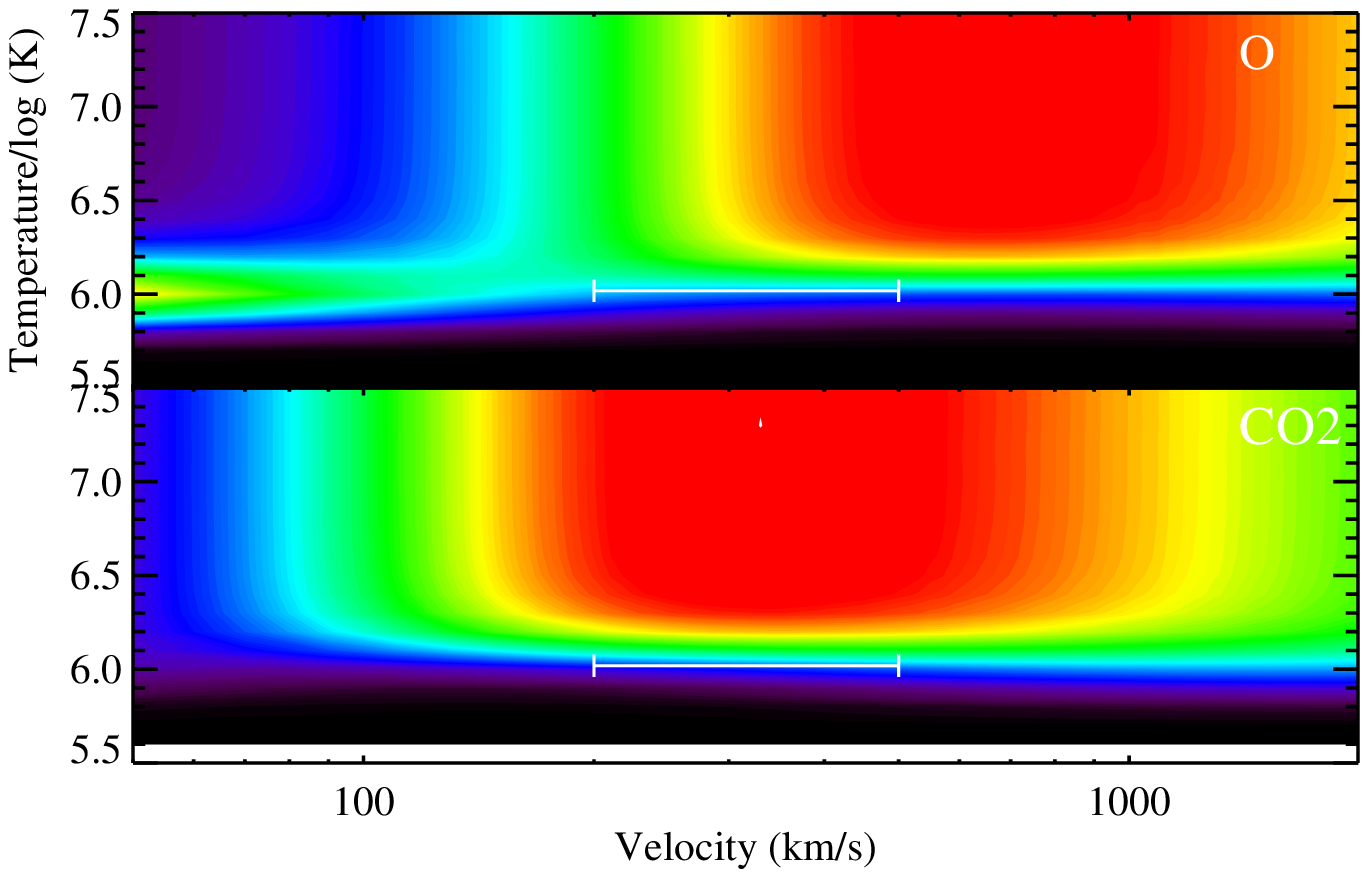}  \includegraphics[width=8.5cm, angle=0]{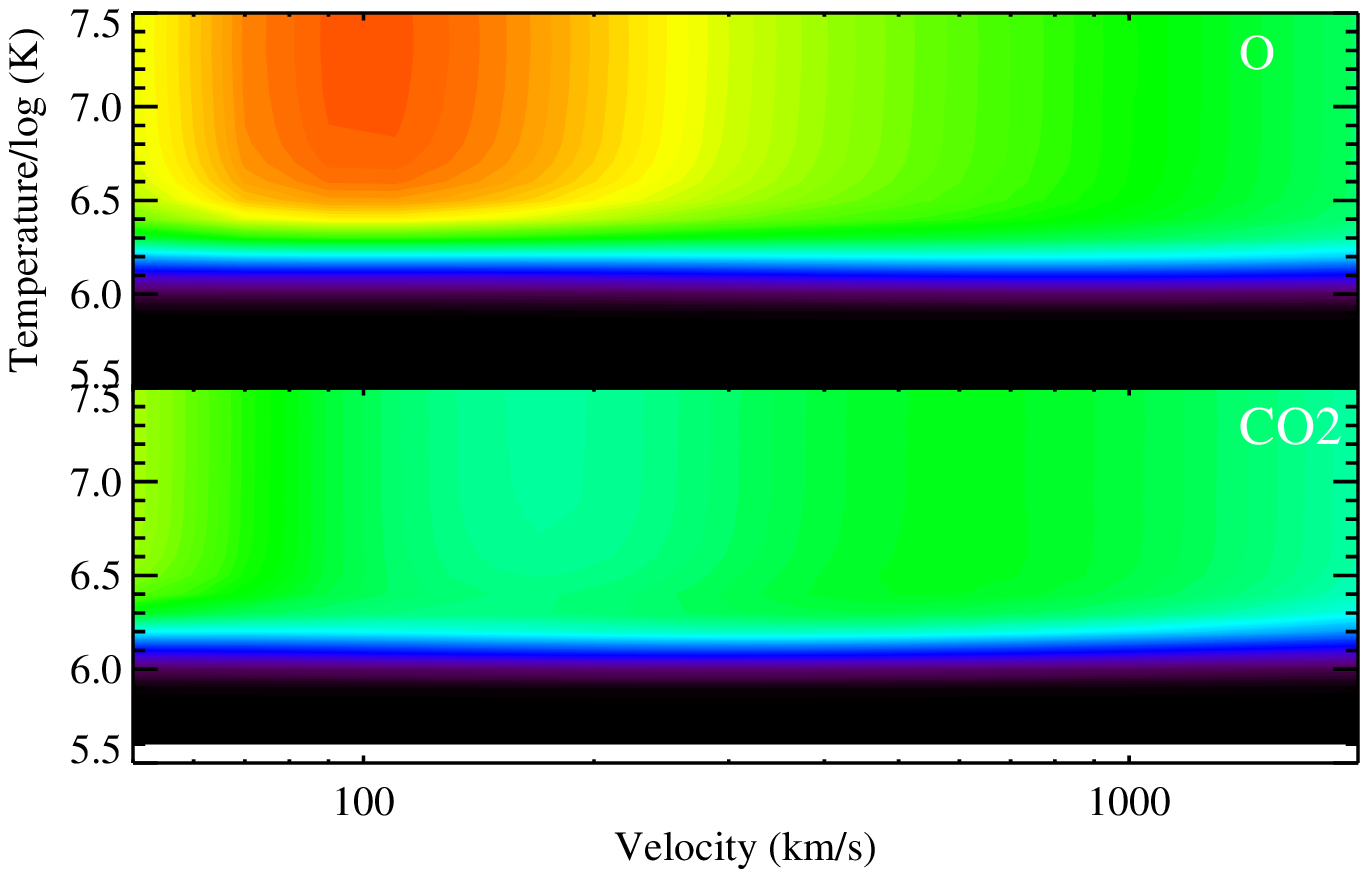}
   \caption{ Contour plots for $\alpha$-values (eV cm$^2$) of carbon (left) and nitrogen (right) in collisions with H, H$_2$, He, O and CO$_2$ in the temperature range log($T_{\rm}$) = 5.0--7.0~K and velocity range of 50--2000 km/s with grid of 0.1 and 20 km/s, respectively. The horizontal white solid lines corresponding to temperature (log$(T)=6.0$ K)  with C$^{5+}$ and C$^{6+}$ abundance of 0.37$\pm$0.03 and 0.13$\pm$0.06 are from ACE data; see Table~1 in \cite{KMC12}. } 
   \label{fig-alpha-cn}
\end{figure*}

\subsection{Evolution of charge stage of solar wind ion}
To examine the validity of the $\alpha$-value that comes from a specified ion abundance at the interaction region, we calculate the evolution of charge stage of solar wind ions of carbon, nitrogen, and oxygen as defined by Eq.~(2). Here, the initial ion abundances are from published literature, for example, \cite{SC00} for oxygen, and \cite{KMC12} for carbon. Because the evolution profile of nitrogen ions are similar to those of carbon and oxygen ions, only bare and H-like carbon and oxygen ions in the solar wind are given. For this we considered the resulting soft x-ray emissions within a photon energy range of 200--1000 eV, see Fig.~\ref{fig-csd-prof}.  This figure shows that the charge stage distribution of solar wind ions does not change over the interaction region between the bow shock (1.57$R_{\rm M}$) and magnetic pileup boundary (1.3$R_{\rm M}$). This indicates that there is no sequential recombination due to charge-exchange when the solar wind ion passes through this interaction region. At an altitude of $\sim$400 km the relative ion abundance of solar wind carbon and oxygen ions starts to change, and reaches a peak ion abundance for H-like ions at the altitude of $\sim$230 km; in other words solar wind ions move slowly, captured H-like ions accumulate and then further sequential recombination take places below $\sim$400 km. By using MAVEN DD2 data for the neutral density~\citep{WCN21}, such rising does not appears, that is ion accumulation and sequential recombination do not take place. This is probably due to the lower oxygen density below 300~km in the MAVEN DD2 data. The evolution behavior of highly charged solar wind ions also indicates that the general fixed $\alpha$-value for a given ion abundance is still valid in the interaction region, but it fails below $\sim$400 km of Mars.
\begin{figure}[th]
   \centering
   \includegraphics[width=8.5cm, angle=0]{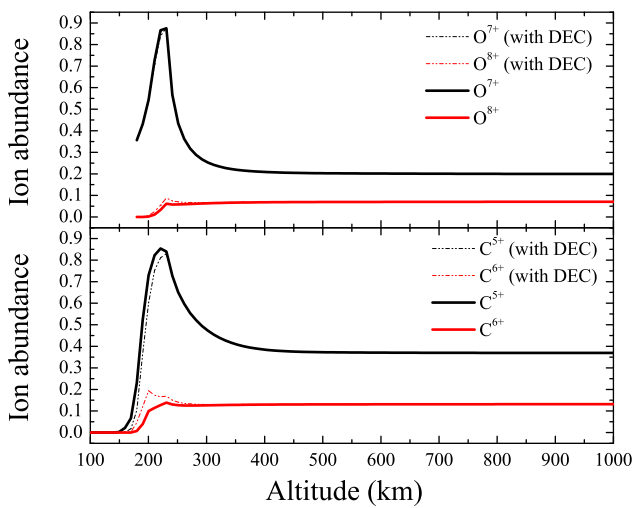}
   \caption{Evolution of charge state of oxygen and carbon ions with initial ion abundance of slow solar wind from the work of  \cite{SC00} for O$^{8+}$/0.07 and O$^{7+}$/0.20, as well as from the work of \cite{KMC12} for C$^{6+}$/0.13 and C$^{5+}$/0.37. Dashed-dot-dot lines refer to calculation with the inclusion of double electron capture (DEC). } 
   \label{fig-csd-prof}
\end{figure}

\subsection{Collisional quenching effect on He-like triplet ratio of O VII}
The triplet-to-single line intensity ratio ($G=\frac{(f+i)}{r}$, here $i, f, r$ is inter-combination, forbidden, and resonance line, respectively) of He-like ions is an important probe for the charge-exchange and/or coronal emissions of astrophysical plasmas, and is used extensively by those who appy astrophysical x-ray spectroscopy with high-resolution~\citep{KTM12,ZWJ14} to study the interface of hot outflows and cold interstellar (or intergalactic) medium. From the high-resolution XMM-Newton observations of Mars, \cite{DLB06} and \cite{KMC12} extracted out these line fluxes for disk and halo regions with $G$ ratios of 0.77$\pm$0.58 (disk) and 5.36$\pm$5.82, respectively. The mean value of the $G$ ratio for the halo is significantly higher than value for the disk. Many charge-exchange emission models have been setup to explain the large $G$ ratio.  \cite{MCL17} and \cite{CML18} listed such velocity-dependent ratios of O~VII for collisions with different neutrals. For comparison, we also plot these ratios in the left panel of Fig.~\ref{fig-ovii-triplet}. The present calculation for collisions with H shows a favorable agreement with the halo mean value and theory from \cite{BCT07}. Due to the low signal-to-noise ratio, the halo $G$ ratio has large error bars (Fig.~\ref{fig-ovii-triplet}). This will be clarified by next-generation X-ray missions, e.g., XRISM~\footnote{https://heasarc.gsfc.nasa.gov/docs/xrism/about/\label{ft-xrism}}, Athena~\footnote{https://www.cosmos.esa.int/web/athena\label{ft-athena}} and HUBS~\footnote{http://hubs.phys.tsinghua.edu.cn/en/index.html\label{ft-hubs}}.
\begin{figure*}[h]
   \centering
   \includegraphics[width=8.5cm, angle=0]{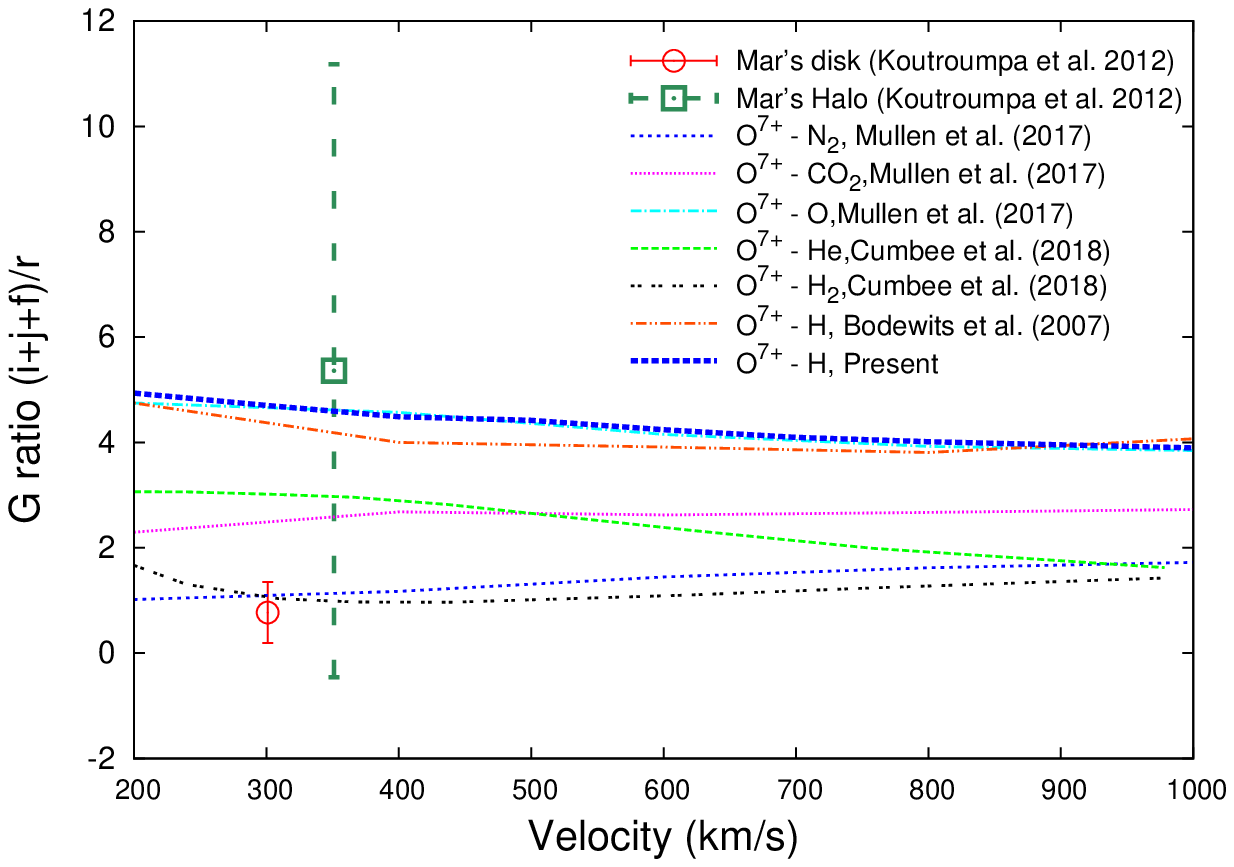} %
   \includegraphics[width=8.5cm, angle=0]{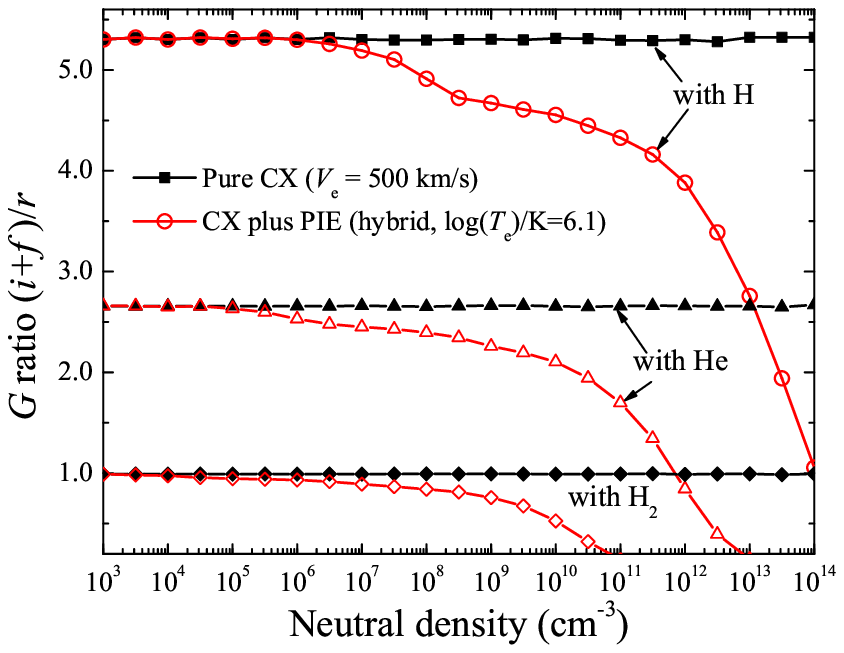}
   \caption{{\it Left:} The $G=\frac{i+f}{r}$ ratio of He-like oxygen. Different charge-exchange (CX) model calculations \citep[and the present work]{MCL17,CML18} are compared along with XMM-Newton observations for Mars's disk and halo from the work of \cite{KMC12}. {\it Right:} The $G$ ratio as the function of neutral density  from a pure charge-exchange model (filled symbols) at typical solar wind velocity of 500 km/s and hybrid one (open symbols) with charge-exchange plus proton/neutral impact excitation (PIE) in collisions with H, H$_2$, and He. Here, the solar wind temperature of log($T_e$)/K=6.1 is based on the previous analysis of the $\alpha$-value of oxygen. } 
   \label{fig-ovii-triplet}
\end{figure*}

For the disk $G$ ratio of 0.77$\pm$0.58 the charge-exchange model from \cite{MCL17} when colliding with nitrogen and  hydrogen gases (N$_2$ and H$_2$) gives ratios ranging from 1.0 to 1.7, which is consistent with disk observations within error bars. \cite{KMC12} suggested this low ratio is due to the quenching collisions with neutrals, that will remove excited electrons from the long-lived metastable states, then suppress the $f$ line intensity. Their qualitative analysis for the collisional effect is clear below 150 km, and they estimated the contribution  from below 150 km to be less than 15\%. In this work we performed a detailed calculation by including excitations from heavy particles in Eq.~(7). However, neutral impact excitation data of highly charged ions are very scarce, even the proton impact excitation of highly charged ions is available for just a few cases. \cite{Sea55,Sea64} presented the theory of proton impact excitation, and applied it to the green coronal line of Fe$^{13+}$for  the $3p_{3/2} \to 3p_{1/2}$ transition. When the excitation energy $\Delta\ E$ is much smaller than the plasma temperature $kT$, the proton excitation rate would be greater than the electron impact by a factor of order $\sqrt{M_p/m}$, $M_p$ being the proton mass and $m$ the electron mass \citep{Dal83}. The excitation energy of O~VII from metastable level (1s2s~$^3S_1$) to 1s2p levels is $\sim$10--16 eV, which is far smaller than the plasma temperature of $\sim$110 eV. Thus, we derived the proton (neutral) impact excitation (PIE) rates from the metastable level of O~VII by using the electron impact excitation rates in the {\sc sasal} database~\citep{LLW14,TKS08}. 

According to Eq.~(7) we calculate the $G$ ratio of O~VII with a pure charge-exchange model and a hybrid one (e.g. CX plus proton impact excitation) in the collisions with H, H$_2$, and He, respectively. From the right panel of Fig.~\ref{fig-ovii-triplet}, we see that the $G$ ratio from the hybrid model starts to deviate from the pure CX one when above the neutral density of 10$^6$ cm$^{-3}$, and becomes more obvious as the neutral density increases. According to the neutral profile presented in Fig.~\ref{fig-mars-neutrals}, the collisional quenching effect appears below the altitude of $\sim$400 km through the collisional excitations by neutral oxygen and/or CO$_2$. Figure~\ref{fig-ovii-triplet} also indicates that the $G$ ratios from the pure CX calculation with different neutrals disperse strongly, for example, 5.3 (H), 2.7 (He), and 0.98 (H$_2$).  However, the disk observation of Mars reported by \cite{DLB06}) covers the photons within 10$^{''}$, that is below altitude of 2180 km. Above $\sim$400 km and below $\sim$2000 km altitude H$_2$, He, and O are dominant components in the Martian atmosphere with density of  $\sim10^4$---2$\times10^5$ cm$^{-3}$. From Fig.~\ref{fig-xraylum} and Fig.~\ref{fig-xray-angle} discussed in next subsection, we can see that the disk observation (white dotted circle) covers more than $\sim$86\% of the emission energies, where the emissivities from H$_2$ and O are comparable to or higher than those from H at lower altitudes. Both their contributions can be up to $\sim$43\% and $\sim$65\%, respectively.  Thus, we suggest the pure CX with H$_2$ and O may be the part of the reason for the low disk observation of 0.77$\pm$0.58. 

By including all the neutral components in Eq.(7) we can further calculate the map of triplet line ratios $G$ with a  pure charge-exchange and a hybrid one as illustrated in Fig.~\ref{fig-gratio-map}. Here, only the hybrid calculation is presented due to both maps  have basically the same distribution except for data below $\sim$200 km which show small differences. Hence, the proton impact excitation may not be the main reason for the low disk observation of 0.77$\pm$0.58. In this figure, the nearly spherical symmetry of the $G$ ratio follows the neutral profiles used in this work.  It is obvious that the\textit{ G} ratio decreases with decreasing altitude, and it reaches a low value of  $\sim$2.8 at altitude of $\sim$700 km. Within the main region of disk observation reported by~\cite{DLB06}, the $G$ ratio  is approximately 2.8--3.4, which is still higher than the disk observation. By using the in-situ MAVEN measurements for neutrals shown in the bottom panel of Fig.~\ref{fig-mars-neutrals}, the $G$ ratio changes slightly to $\sim$2.2--3.0. This decreasing trend of the $G$-ratio indicates that the contributions from H$_2$ and O steadily increase and become significant, but still not absolutely dominant.  This is consistent with the x-ray luminosity profiles shown in Fig.~\ref{fig-xray-angle}, which will be discussed in the next subsection. When the solar wind bulk velocity is used, the $G$-ratio can reach a low value of $\sim$0.9 at the altitude of $\sim$400 km. The higher cross-section below 100 km/s in the collision with H$_2$ (see Fig.~\ref{fig-cs-o8-neus}) and the low bulk velocity ($\le$100 km/s, see Fig.~\ref{fig-swden}), is the reason for the low calculated $G$-ratio of $\sim$0.9. Therefore, we suggest that the charge-exchange with H$_2$ gas may still be the possible reason for the low disk observation. Such dependence of the $G$ ratio on the altitude can be explored by future deep observations with next generation x-ray missions with high efficiency (e.g. XRISM~$^{\ref{ft-xrism}}$, Athena~$^{\ref{ft-athena}}$ and HUBS~$^{\ref{ft-hubs}}$). We also notice that there is a tail-like feature for the $G$-ratio by using the bulk velocity of the solar wind, that is consistent with the tail-like feature in the bulk velocity map shown in Fig.~\ref{fig-swden}. Then we think this tail-like feature in the $G$-ratio is resultant from the low bulk velocities and relative higher cross-sections at low collisional velocities with the multiple-electron neutrals (e.g. H$_2$, He and O) as shown in Fig.~\ref{fig-cs-o8-neus}. By considering the observed $G$ ratio derived from the observed flux in the line of sight, the $G$ ratio map is weighted by projected flux in different direction of line of sight (LOS) discussed in Sect. 3.5, then we obtain the expected $G$ ratio of $\sim$1.6--1.8 in the disk observation, see Table~\ref{tab-flux}. That is slightly higher than the disk observation of 0.77$\pm$0.58. It should be noted that the present charge exchange cross sections are the best ones available, but not the best ones  qualified  to use for high-resolution spectroscopy; these need more  elaborate benchmark measurements for the $nl-$selective cross section for its extensive application in the near future.
\begin{figure*}[h]
   \centering
    \includegraphics[width=8.cm, angle=0]{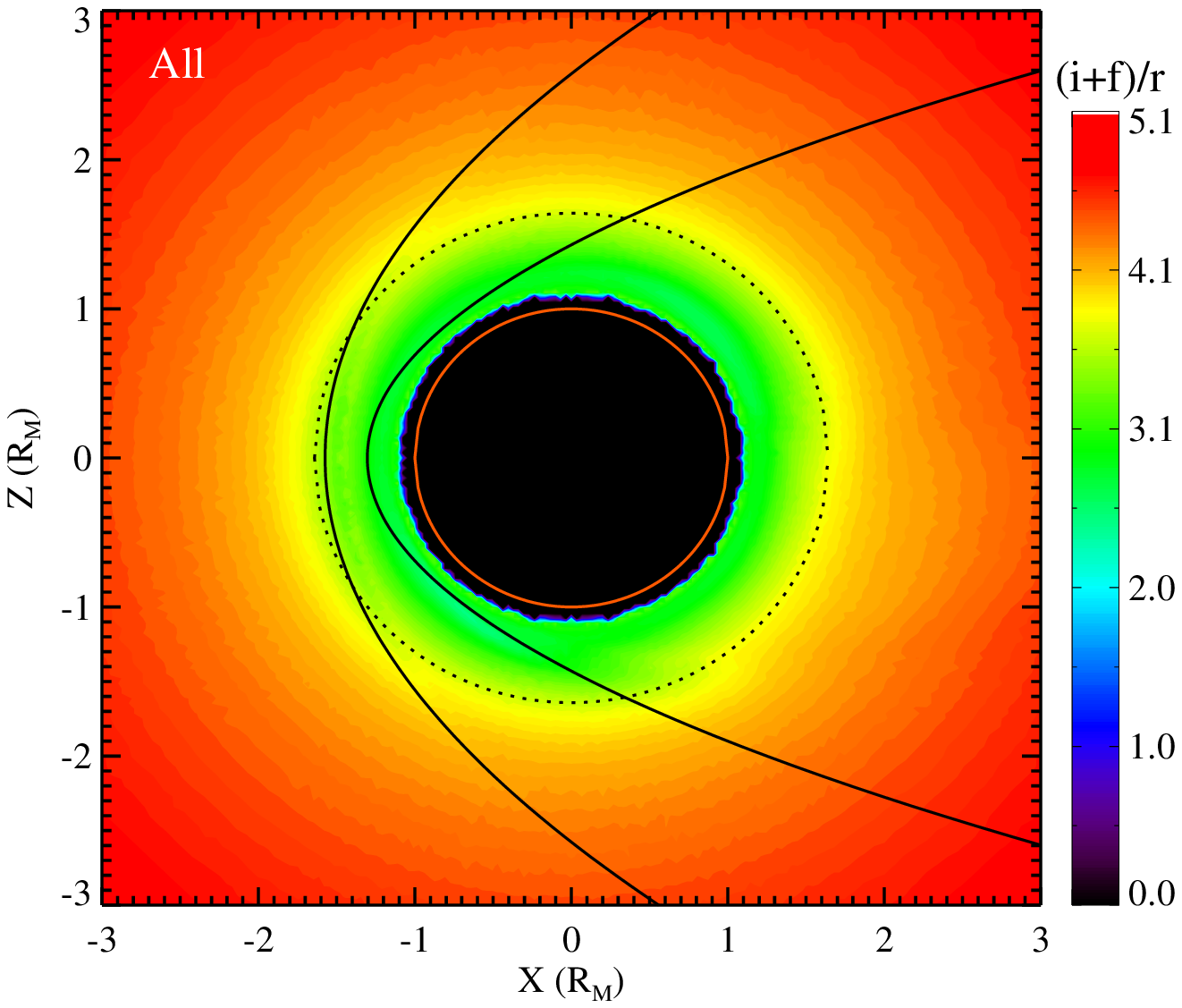} 
    \includegraphics[width=8.cm, angle=0]{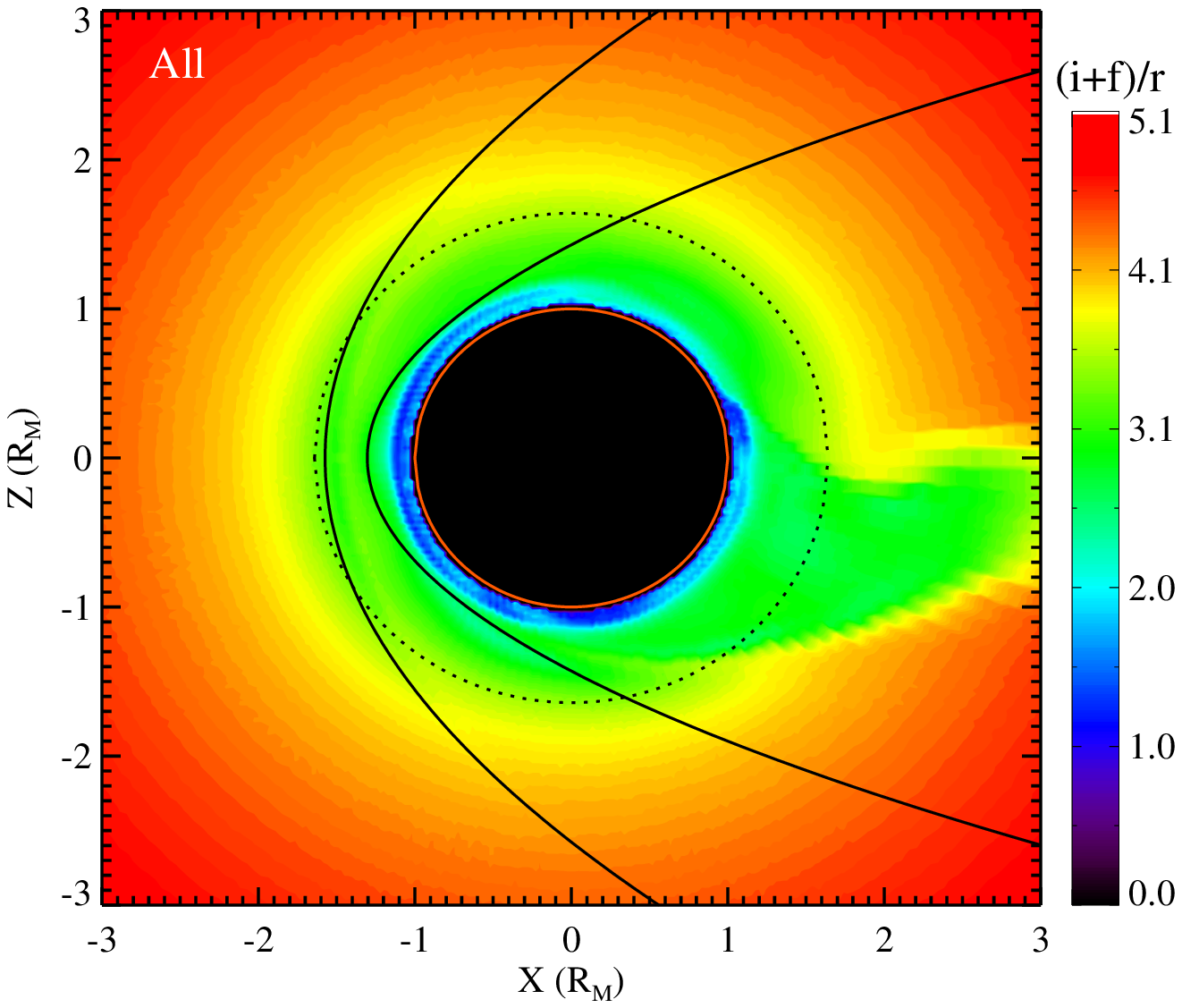} 
   \caption{Map of $G=\frac{i+f}{r}$ ratio of He-like oxygen with charge-exchange and proton impact excitation (namely, the hybrid model) for all neutrals in cases of average collisional velocity $v$ ({\it left}) and solar wind bulk velocity $v_{\rm sw}$ ({\it right}). Solid curves refer to the bow shock and magnetic pileup boundary from the work of \cite{NLN20}. Black dotted and red solid circles indicate regions of disk observation in \cite{DLB06} and Mars's position, respectively. {\it Note: Both the pure and hybrid calculation are basically the same except for values below $\sim$200 km.}} 
   \label{fig-gratio-map}
\end{figure*}

\subsection{Contribution of different neutrals on x-ray emissivity distribution in the XZ plane}
In view of the different neutral profiles as shown in Fig.~\ref{fig-mars-neutrals} and different charge-exchange cross-sections as shown in Fig.~\ref{fig-cs-o8-neus}, we begin by investigating the contribution from different neutrals to the x-ray  emissivity distribution in the XZ-plane (Fig.~\ref{fig-xraylum}). The spatial distribution of x-ray emissivities shows obvious differences in collisions with different neutrals.  On the whole the x-ray emissivity distribution for the collisions with H shows an obvious bow shock in the XZ-plane, consistent with that derived from MAVEN spacecraft data \citep[solid white curves]{NLN20} and the density map from the MHD simulation.  Weak x-ray emissions extend toward the solar direction in the longer region and toward the far region in magnetosheath. However, the bright x-ray emissions in the collisions with H$_2$ (with peak value of 1.5$\times10^{-14}$ erg~cm$^{-3}$s$^{-1}$) concentrate in the region near the magnetic pileup boundary with less extension toward the magnetosheath, and are higher than those from the H collision between $\sim$1.15--1.32$R_{\rm M}$ (Fig.~\ref{fig-xray-angle}). The contributions of the He collision are mainly below $\sim$1.17$R_{\rm M}$ with emissivity values much smaller than those of the H and H$_2$ collision. Since the oxygen and carbon-dioxide become the dominant components of the Martian atmosphere below the altitude of $\sim1.16R_{\rm M}$ (550~km), the main x-ray contribution is from the collision with O with emissivity values  comparable to those of H collisions, as shown by the dark-blue curve in Fig.~\ref{fig-xray-angle}. For a clear inspection we further present the profile of x-ray emissivities at four different directions (0$^{\circ}$, 30$^{\circ}$, 60$^{\circ}$ and 90$^{\circ}$) relative to the  Sun-to-Mars line (Fig.~\ref{fig-xray-angle}). Basically, the contributors to x-ray emissivity profiles follow the neutral profiles in the Martian environment with some difference in detail.
\begin{figure*}[th]
   \centering
   \includegraphics[width=5.7cm, angle=0]{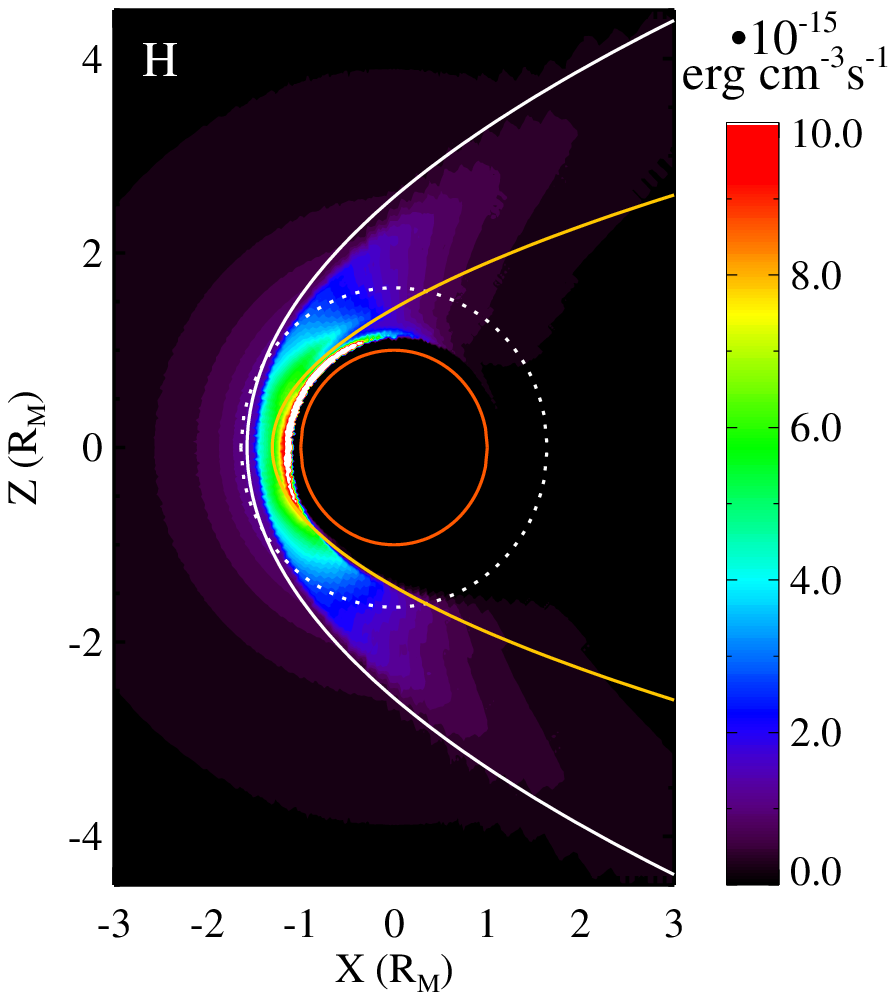} 
   \includegraphics[width=5.7cm, angle=0]{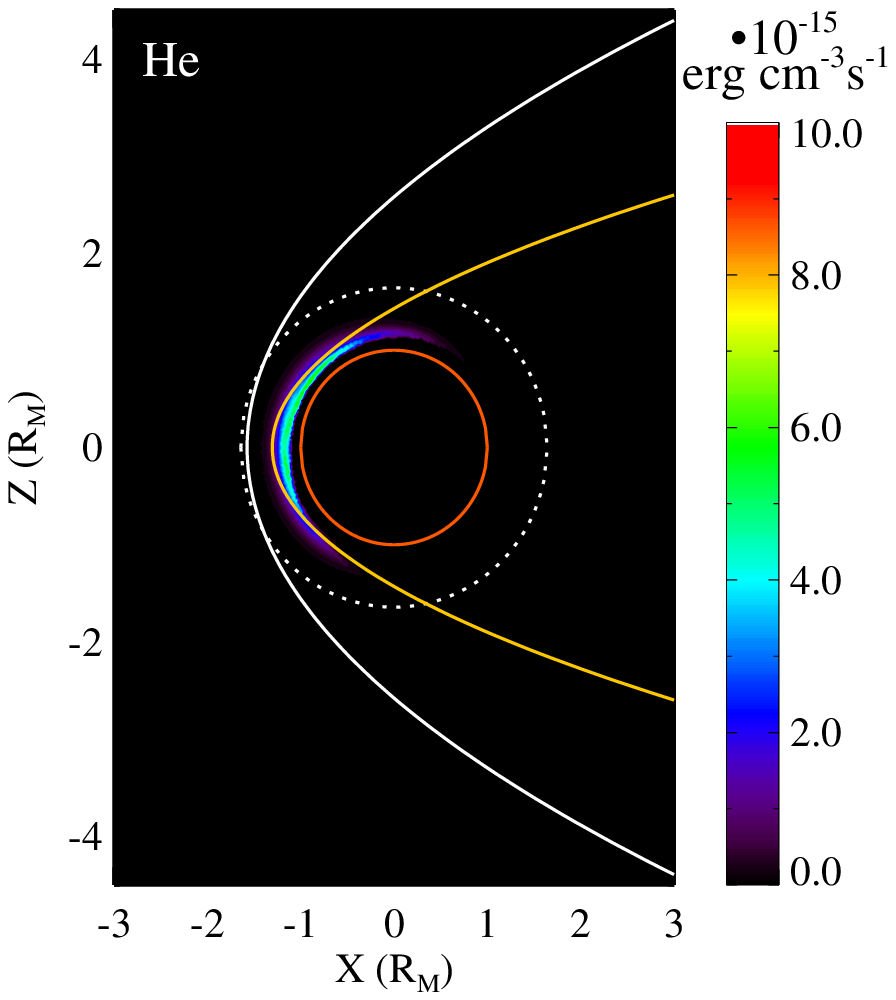} 
   \includegraphics[width=5.7cm, angle=0]{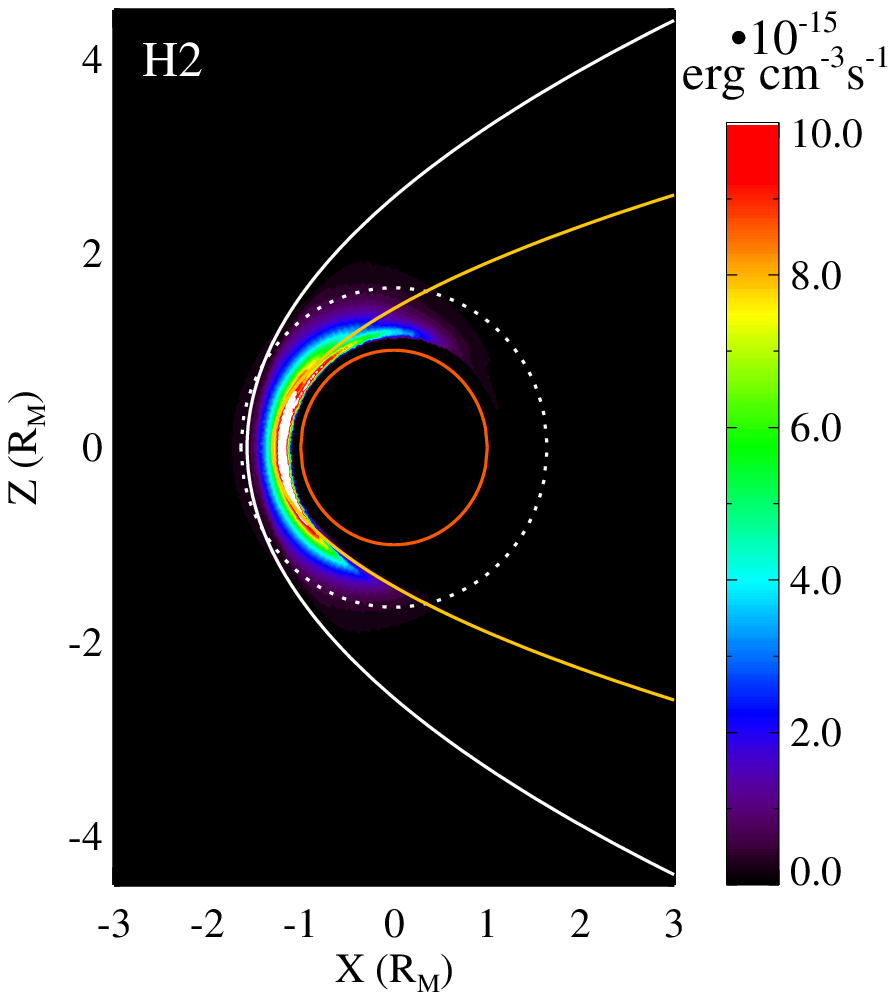} 
     \includegraphics[width=5.7cm, angle=0]{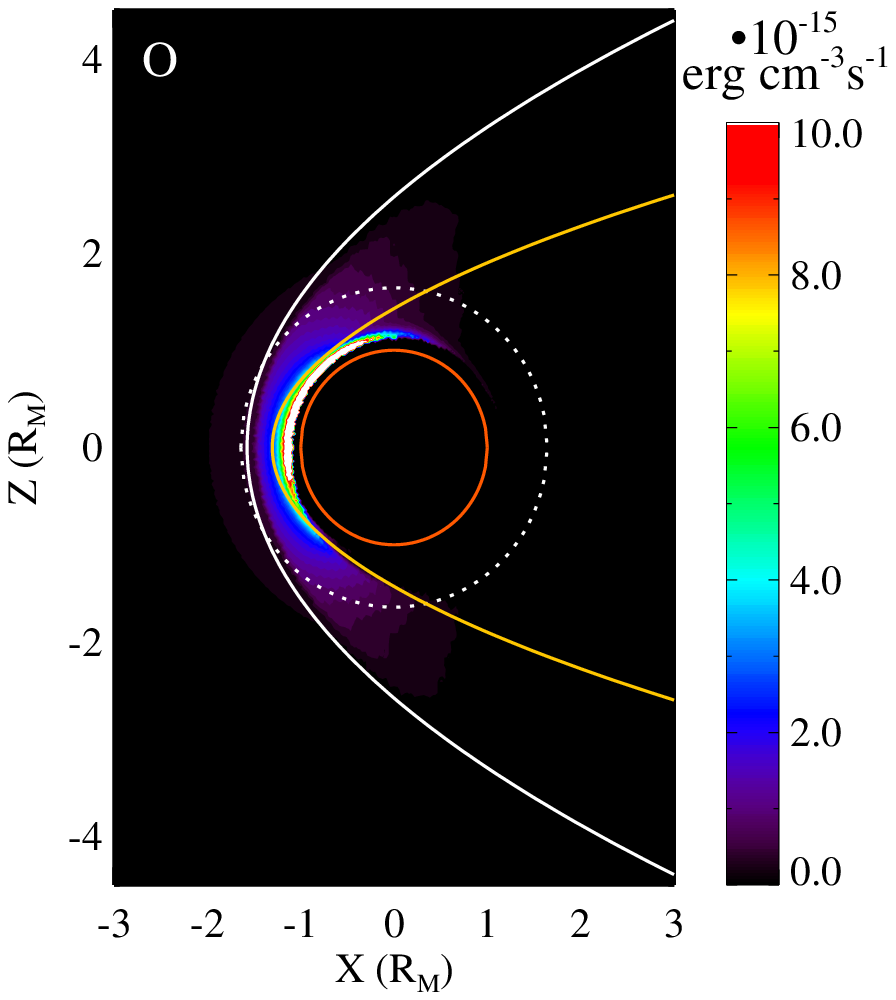} 
     \includegraphics[width=5.7cm, angle=0]{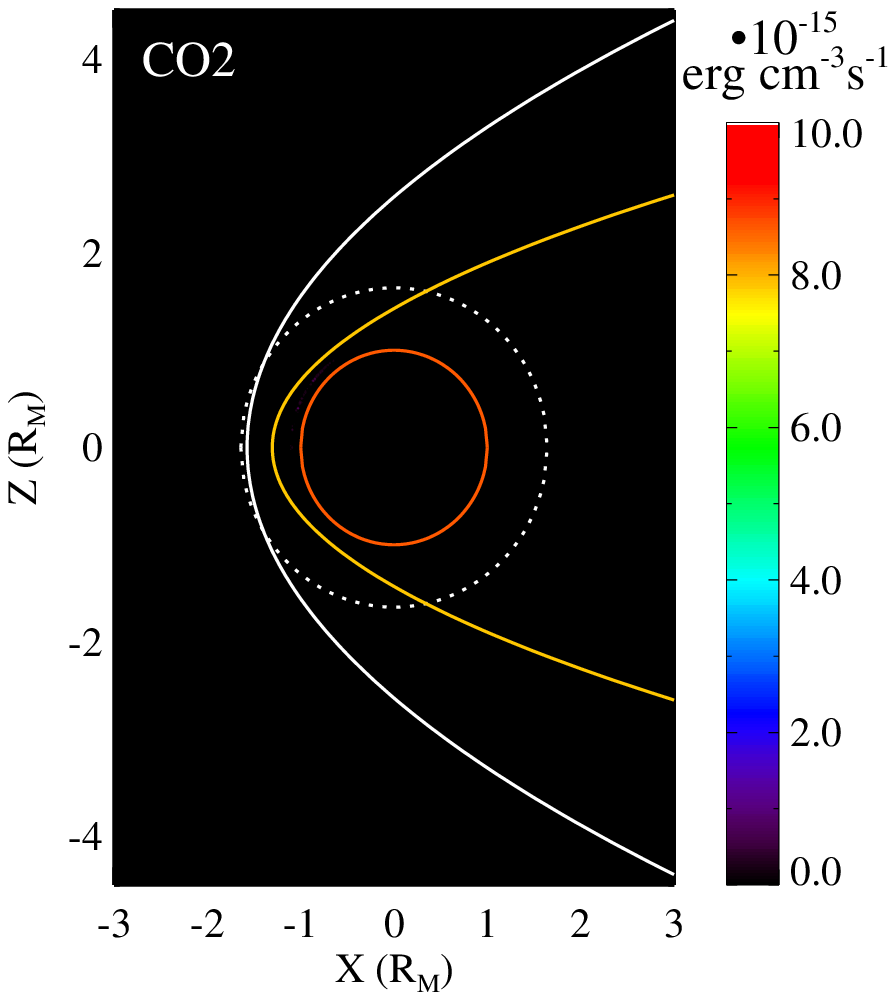} 
      \includegraphics[width=5.7cm, angle=0]{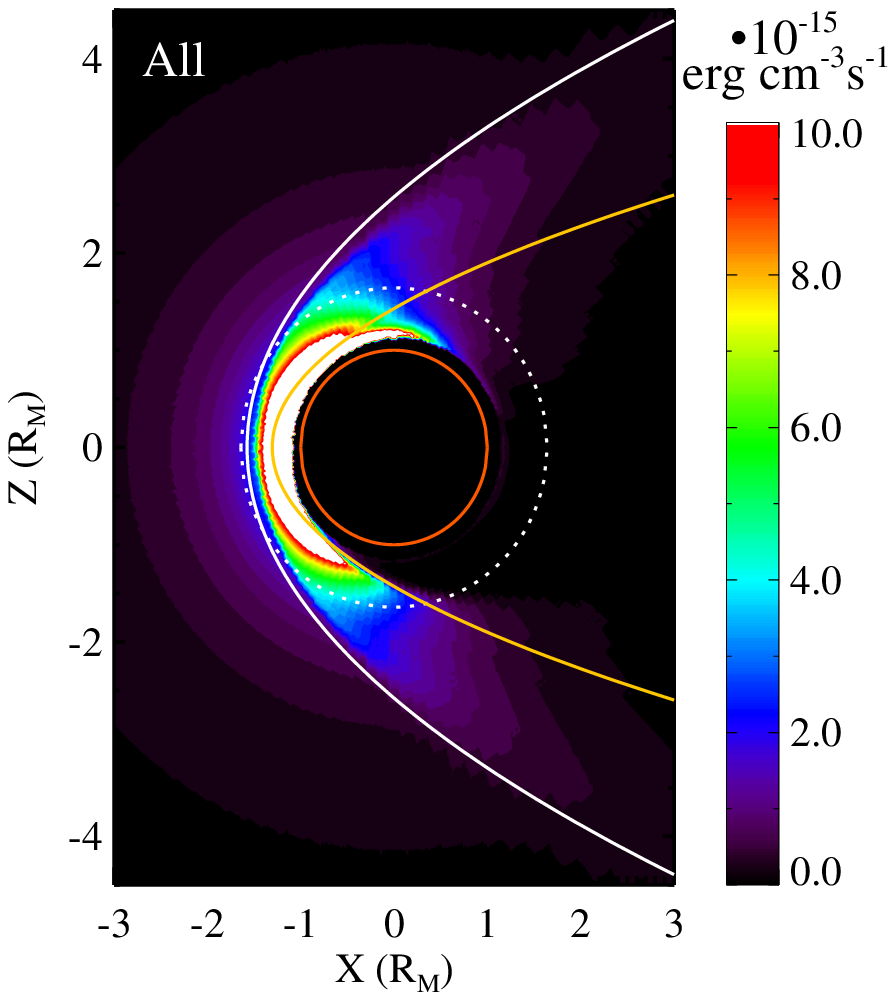}
   \caption{X-ray emissivity (with photon energies of 0.2---1.0~keV) map in XZ-plane of highly charged ions (O) in collisions with neutrals of H, H$_2$, He, O, CO$_2$ and all around the Martian atmosphere. The spatial units are in terms of the martian radius $R_{\rm M}$. Solid curves refer to the bow shock and magnetic pileup boundary from the work of \cite{NLN20}. White dotted and red solid circles indicate regions of disk observation in \cite{DLB06} and Mars's position, respectively. {\it Notes:} For the collision with CO$_2$, the emissivity is too weak to be manifested, see Fig.~\ref{fig-xray-angle} for details.
   } 
   \label{fig-xraylum}
\end{figure*}

\begin{figure}[h]
   \centering
   \includegraphics[width=9.0cm, angle=0]{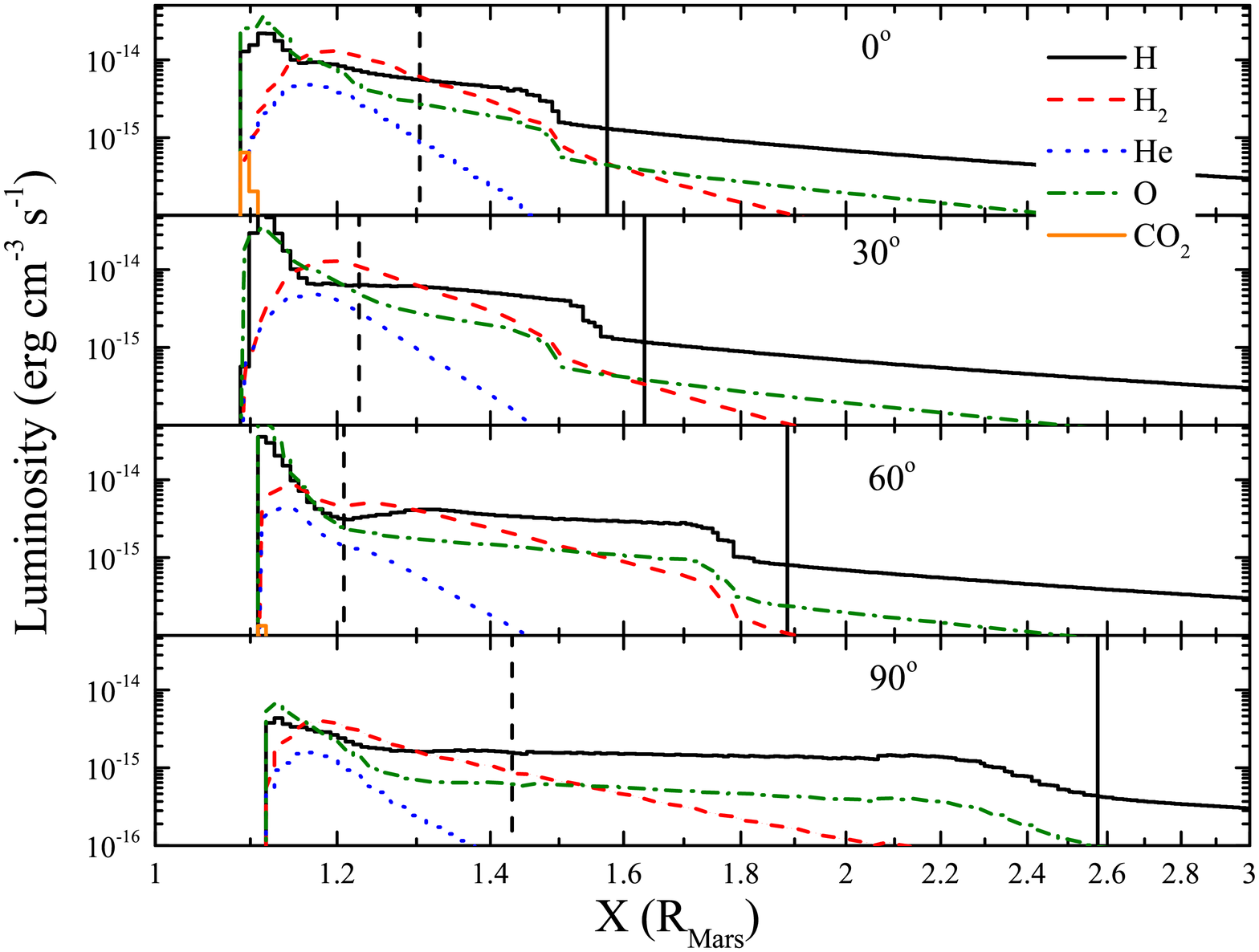} %
   \caption{X-ray emissivity profile of oxygen in XZ-plane at different directions (0$^{\circ}$, 30$^{\circ}$, 60$^{\circ}$ and 90$^{\circ}$) relative to solar-Mars center line. Vertical solid and dashed lines mark the positions of bow shock and magnet pileup boundary from the fitting formula based upon MAVEN satellite data ~\citep{NLN20} .
   } 
   \label{fig-xray-angle}
\end{figure}

By considering the consistency of x-ray emissivity distribution with the structure of solar wind interaction with the Martian exosphere, the present work confirms again that x-rays can be used to probe the global structure of solar wind interaction with planets~\citep{SCK04,SCC09,SWS19}. The different contribution profiles and line features  in x-rays with different neutrals suggest that x-ray spectroscopy can  probe the neutral components with future deep observations with spatial and energy resolutions that can be achieved by next generation x-ray missions, such as XRISM~$^{\ref{ft-xrism}}$,  Athena~$^{\ref{ft-athena}}$ and HUBS~$^{\ref{ft-hubs}}$.

\subsection{X-ray morphology of Mars}
The same procedure is used for carbon and nitrogen emissions with ion fractions of 0.13 (C$^{6+}$)\footnote{From real-time measurement of ACE-SWICS, see Table 1 in the work of \cite{KMC12} \label{ft-k12}}, 0.37 (C$^{5+}$)$^{\ref{ft-k12}}$, 0.006 (N$^{7+}$), and 0.058 (N$^{6+}$). We calculate the x-ray emissivities in the energy range between 200---1000~eV for the three-dimensional distribution of carbon, nitrogen, and oxygen ions within 8.3$R_{\rm M}$ (55$^{''}$) of Mars. Since there are not emission lines from  captured H- and He-like carbon, nitrogen and oxygen ions in the ranges of 200--300~eV and 900--1000~eV \citep[see Table~1 and Fig.~10 there, respectively]{KLK06,LZW21}, the present energy range is consistent with that (300--900~eV) in the XMM-Newton observation and previous works \citep{KMC12}. We obtain the x-ray morphology and total  x-ray luminosity around Mars (Fig.~\ref{fig-lum-tot}) by integrating along the $y$-direction (Martian motion) and summing in this projected plane as following:
\begin{eqnarray}
L(xz) = \int \sum_{ele,q+} P^{q+}(xyz) dy, \\
L_{tot} = \int \sum_{ele,q+} P^{q+}(xyz) dV,
\end{eqnarray}
where $L(xz)$ refers to the projected emission flux (in the unit of erg cm$^{-2}$~s$^{-1}$) in $y$-direction with integration range between -8.3$R_{\rm M}$ and $+$8.3$R_{\rm M}$, yet those emissions from the $y< 0$ region with $\sqrt{x^2+z^2}\leq R_{M}$ are blocked by Mars in this projection plot. The summations $\sum_{ele,q+}$ are for above listed elements, as well as H- and He-like captured ions that emits x-ray photons.  $L_{tot}$ is the total SWCX luminosity by integrating the emissivity $\sum_{ele,q+} P^{q+}(xyz)$ within a cubic box with the size of 16.6~$R_{\rm M}$ (or $\pm8.3R_{\rm M}$). The emission rate $P^{q+}(xyz)$ or $P^{q+}(r,\theta,\phi)$ of one charged ion is defined by Eq.(1) in the previous model section.

The resulting total SWCX x-ray luminosity is 6.55~MW (O: 3.01, C: 2.69 and N: 0.85~MW, respectively) in this work, showing a better agreement with the XMM-Newton observation of 12.8$\pm$1.4 than previous predictions.  \cite{KMC12}  made some estimations for the additional x-ray luminosity, e.g., larger simulation box  ($\sim$8~$R_{\rm M}$) with an additional 5\%,  average abundance of solar wind  which is three times higher than their usual value, with He and H$_2$ contributions in the disk region. After these additional contributions their simulated x-ray luminosity can reach between 1.2 and 2.0~MW. They also pointed out a halo coronal mass ejection (CME) event on 2003/11/18 as perhaps a possible explanation. A strong ion flux on average 18 times real-time values in their simulation could yield a total luminosity of $\sim$6.3~MW, that is in better agreement with, but still quite lower than, the observed value of 12.8$\pm$1.4~MW.  

In this observed luminosity, assumptions of isotropic emission and optically thin were used by multiplying the observed fluxes of emission lines with $4\pi\Delta^2$ (here $\Delta=0.77$~AU is the distance between Earth and Mars)~\citep{DLB06,KMC12}. Contribution from fluorescent scattering of solar x-rays has not be included~\citep{DLB06}. By using the observed fluxes of the fluorescence lines from $1\pi_g \to 1s$ and $3\sigma_u \to 1s$ transitions of CO$_2$ around $\sim$525~eV, the fluorescent luminosity was derived to be 3.4$\pm$1.4~MW by \cite{DLB06}, being approximately 27\% of the total SWCX luminosity of 12.8$\pm$1.4~MW.  

We further calculate total SWCX luminosity contributions at different layers (with a step of 0.5~$R_{\rm M}$) by using the oxygen emissivity with all listed neutrals, see Fig.~\ref{fig-totlum-dist}. The contribution from the interaction region ($<2R_{\rm M}$) is the largest emission source ($\sim21\%$) with the minimum emitting volume. While the emissivities shown in Fig.~\ref{fig-xraylum} become smaller with increasing altitude, the emitting volume increases as $R^2_{\rm M}$, resulting in the luminosity contributions to increase again after $\sim4.5~R_{\rm M}$.  The luminosity contribution between 43$^{''}$ ($\sim7R_{\rm M}$) and 50$^{''}$ radius ($\sim8R_{\rm M}$) is $\sim16\%$, being higher than the crude estimation (5\%) of \cite{KMC12}. This illustrates that the uncertainty of neutrals at high altitude has a non-negligible effect on the total luminosity. However, the absence of in-situ measurement for the hydrogen density at higher altitudes limits the examination for the present calculation. 

By using a different group of neutral densities, e.g. in-situ MAVEN measurement at the orbit period of DD2~\citep{SYB22,WCN21}, we re-calculate the total CXE luminosity to be 5.93~MW being smaller than the former calculation by 10\%. Although there are large differences for the neutral densities at low altitude shown in Fig.~\ref{fig-mars-neutrals}, the resultant total SWCX luminosity does not change a lot. Figure~\ref{fig-mars-neutrals} illustrates that SWCX emission is dominantly from the collision with hydrogen at $\gtrsim2R_{\rm M}$, where the hydrogen density is very close between the fitting to MAVEN DD2 data and the fitting given by Eq.~(3). The small difference of the calculated luminosities indirectly indicates that the emission from distant halo regions with large volume around Mars plays an non-negligible role on the observed luminosity, that is consistent with the discussion for the radial distribution of the luminosity, as shown in Fig.~\ref{fig-totlum-dist}.

By using the in-situ solar wind data from ACE Science Center Level 2 database~\footnote{https://www.swpc.noaa.gov/products/ace-real-time-solar-wind\label{ft-ace}}, we estimate the time delay {\bf ($dt$)} of the solar wind to be about $\sim$45--53 hours between the L1 point (ACE position) and Mars by using $dt= \Delta/v_{\rm sw}$ with $v_{\rm sw}$ of $\sim$600--710~km/s and  $\Delta$=0.77~AU \citep{KMC12}. This corresponds approximately to the window between 2003/11/18 05:00 UT and 2003/11/19 10:00 UT for the solar wind event around Mars, during the XMM-Newton observation \citep{DLB06}, see Fig.~\ref{fig-ace-data}. The solar wind velocity decreases from $\sim$710 to 530~km/s, yet the density holds basically a constant value of 2.7~cm$^{-3}$ till 2003/11/19 00:00 UT, then increases to $\sim$5.0~cm$^{-3}$ at the end of above window. By considering the difference between the in-situ ACE measurements and the values used in this work, the total SWCX luminosity varies in the range of 6.28--8.68~MW. When the solar wind density is further scaled to Mars' heliocentric distance (1.43~AU) by $1/r^2$, the total luminosity decreases to $\sim$3.1--4.3~MW.  During this period, the solar wind state changes slightly at Mars, see gray shadow region in the third panel of Fig.~\ref{fig-ace-data}. By using the ACE measured oxygen abundance and the charge state \citep{BCX21}, we further calculate the $\alpha$-value of oxygen, that shows a mean value of (1.4--6.6)$\times10^{-16}$~eV~cm$^2$ from the beginning till the end of the event window at Mars, that is slightly lower than the value of 7.3$\times10^{-16}$~eV~cm$^2$ adopted in our work, see the discussion at Sect.3.1. That is the resultant total luminosity should be decreased again slightly.
\begin{figure}[th]
   \centering
   \includegraphics[height=6.6cm, angle=0]{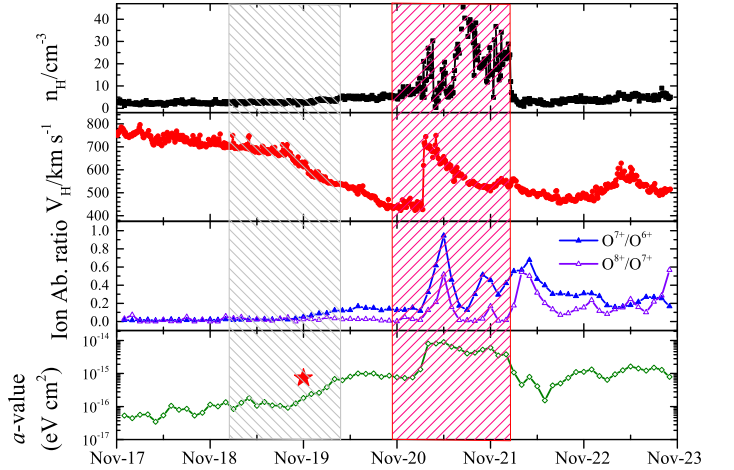}
   \caption{Solar wind properties (proton density $n_{\rm H}$, bulk velocity $v_{\rm sw}$, ion abundance ratio of O$^{7+}$/O$^{6+}$ and O$^{8+}$/O$^{7+}$) as measured by ACE during 2003/11/17--23,  and $\alpha$-value of oxygen in the collision with H (bottom) at a velocity of 500~km/s by using the in-situ ACE data. Red shadow region marks the window of XMM-Newton observation, while gray shadow region marks the window at Mars. The filled star symbol in the bottom panel is the $\alpha$-value of 7.3$\times10^{-16}$~eV~cm$^2$ discussed in Sect.3.1 with O$^{7+,8+}$ abundances of 0.28 and 0.05, respectively. {\it Notes: $n_{\rm H}$ and $v_{\rm sw}$ are binned by 12-min, while ion abundance ratio is binned by 2-hr from the ACE Science Center Level 2 database.}   } 
   \label{fig-ace-data}
\end{figure}

Another possible reason is the assumption of isotropic emission used to derive the observed luminosity by \cite{DLB06}. The projected x-ray flux shown in the left panel of Fig.~\ref{fig-lum-tot} illustrates that there is an obvious non-isotropic feature for the dayside and nightside. By considering the observed flux from a specified phase angle ($\phi\approx40^{\circ}$) adopted to derive the luminosity, we calculate the mean value for the projected x-ray emission flux within the disk region (15$^{''}$ radius) at different phase angles, see Table~\ref{tab-flux}. It shows that the mean projected flux varies within 4.5--5.7$\times10^{-6}$~erg cm$^{-2}$ s$^{-1}$. Then we suggested the observed luminosity of 12.8$\pm$1.4  might be overestimated by $\sim$20\%.
\begin{table}
\centering
\caption{Mean projected x-ray emission fluxes within disk region at different phases.} \label{tab-flux}
\begin{tabular}{lccc}
\hline\hline
Phase  & $G$ ratio & Projected flux  \\ 
 angle    &           & $10^{-6}$~{\rm erg~cm$^{-2}$~s$^{-1}$} \\\hline
90$^{\circ}$ & 1.75 &  5.60 \\ 
60$^{\circ}$ & 1.63 &  4.52 \\ 
40$^{\circ}$ & 1.75 &  5.61 \\ 
0$^{\circ}$  & 1.68 &  4.79 \\ 
120$^{\circ}$& 1.64 &  4.49 \\
140$^{\circ}$& 1.74 &  5.66 \\
180$^{\circ}$& 1.68 &  4.76 \\
\hline
\end{tabular}
\end{table}

In summary, the total SWCX luminosity is closely related to the solar wind condition and the planetary environment, that can be used to study the neutral density when the real-time information is available for the solar wind.

\begin{figure*}[th]
   \centering
   \includegraphics[height=7.cm, angle=0]{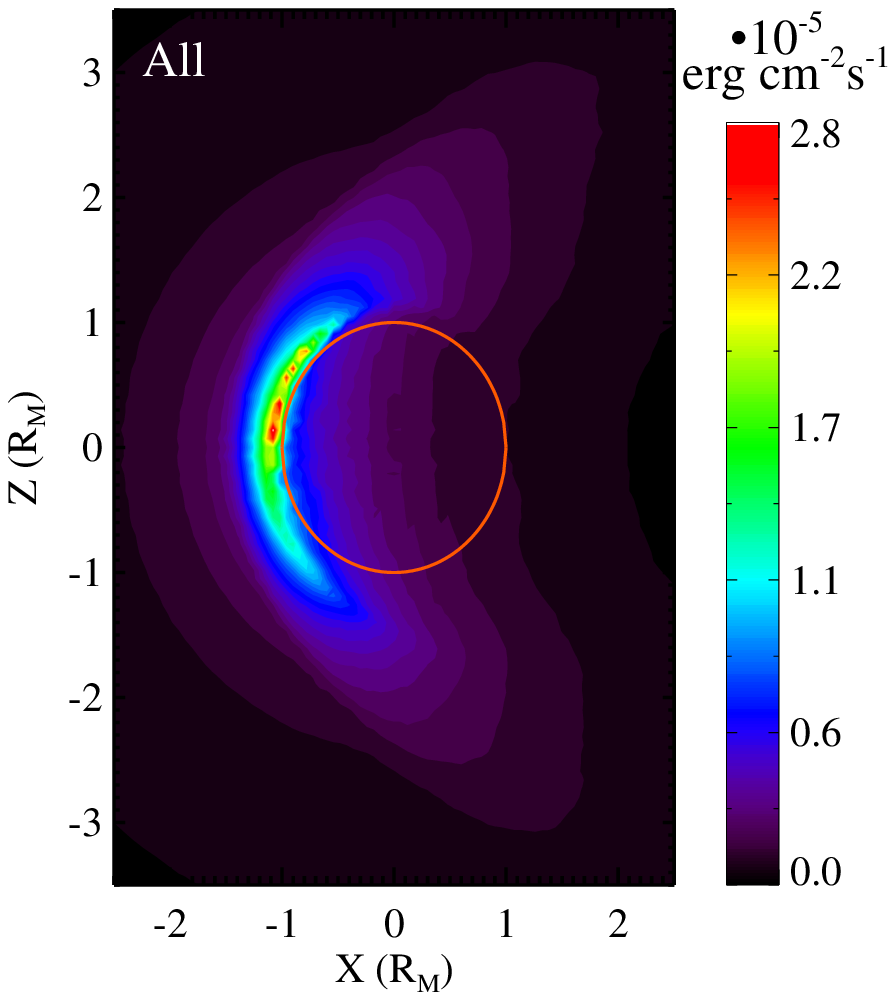}
  \includegraphics[height=7.cm, angle=0]{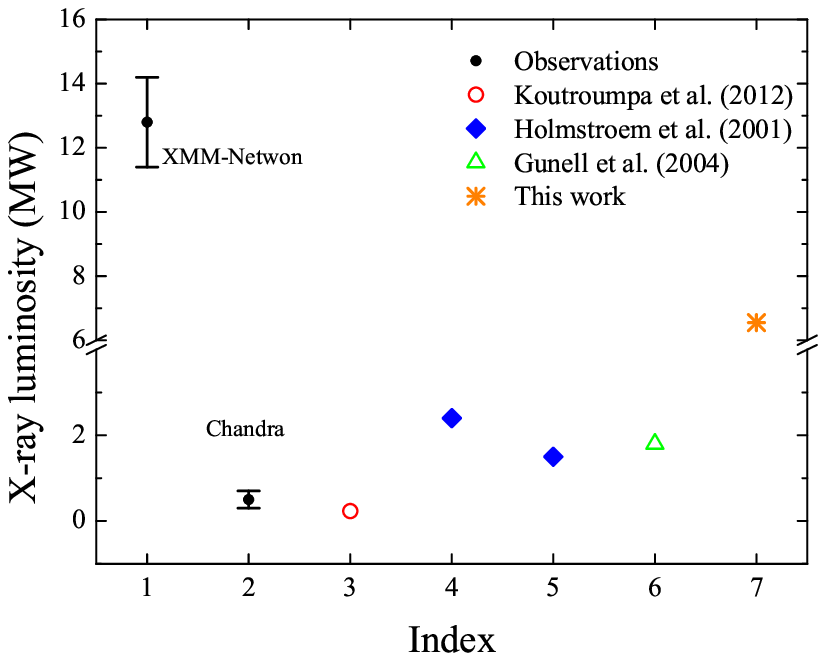}
   \caption{{\it Left}: Projection in Martian motion direction ($y$)  of three-dimensional x-ray {\bf emissivity} distribution of solar wind oxygen ions with all neutrals.  {\it Right}: Total luminosity around Mars from observations of XMM-Newtron \citep{DLB06} and Chandra \citep{Den02}, and from previous predictions and this work. } 
   \label{fig-lum-tot}
\end{figure*}

\begin{figure}[th]
   \centering
   \includegraphics[width=9.cm, angle=0]{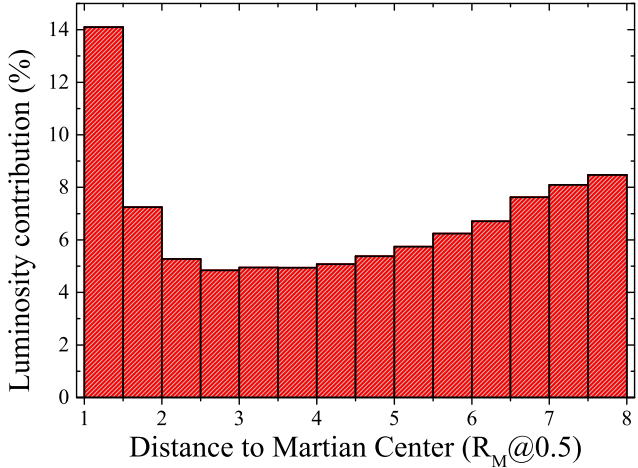}
   \caption{Radial distribution of the total luminosity in the Martian exosphere with an grid step of 0.5$R_{\rm M}$.} 
   \label{fig-totlum-dist}
\end{figure}

\section{Summary and conclusion}
In this study,  using the newest charge-exchange cross section in Kronos v3.1 and experimental measurements, we calculate the emission factor $\alpha$-value of carbon, nitrogen, and oxygen with different neutrals (H, He, H$_2$, O, and CO$_2$) in the Martian environment  over wide temperature and velocity ranges. The $\alpha$-value is highly variable over a temperature range of log($T_e$)/K=5.9---6.4 for oxygen, which shows an obvious dependence on velocity below 200--300 km/s. Both the MHD simulation and in-situ MAVEN measurements reveal that the bulk velocity of solar wind decreases to below 100~km/s after the bow shock. Then the general single $\alpha$-value is not valid again. Overall, the $\alpha$-value in collisions with O and CO$_2$ is higher than with others, e.g., H and He. The present $\alpha$-value of oxygen agrees well with previous reports in the collisions with H at the temperature of log($T_e$)/K=6.1 and typical solar wind velocity of 300--600~km/s.  The evolution of charge stage of solar wind ions shows that there is not a sequential recombination from charge-exchange across the interaction region; H-like ion pileups and sequential recombination appear below the altitude of 400~km. This indicates again that the general fixed $\alpha$-value is not valid below this altitude.

By considering the excitation energy ($\sim$10--16~eV) from the metastable level to higher $1s2p$ levels, we obtain the proton impact excitation cross sections from electron impact excitation data based on the theory of \cite{Sea55}. Then we incorporate them into a sophisticated hybrid emission model. Furthermore, the anonymous low disk $G=\frac{i+f}{r}$ ratio (0.77$\pm$0.58) was explored, and can be directly explained  by the collisional quenching effect due to proton/neutral collisions.  However, the quenching contribution is small for the disk observation and only appears below 400~km. Hence, we suggest that charge-exchange with H$_2$ and N$_2$ may be the most likely reason for this low mean $G$-ratio with large error-bars.  

We also presented x-ray emissivity maps from solar wind ions impinging on different neutrals in the Martian exosphere, which is in accordance with bow shock derived from in-situ MAVEN solar wind ion density and velocity mapping. The contributions from different collisional neutrals are explored, which are shown to differ from each other.  The resulting total x-ray luminosity of 6.55~MW shows a better agreement with the XMM-Newton observation of 12.8$\pm$1.4~MW than previous ones. Its dependences on solar wind variation and neutral density profile around Mars are discussed. 

We present a detailed study for x-rays due to charge-exchange around Mars. This reconfirms  that x-rays represent a good remote sensor for the global interaction of solar wind with a planetary atmosphere. This study illustrates an example of charge-exchange emissions in space physics, and shows a requirement for benchmarks for data of $nl-$selective velocity-dependent charge-exchange cross-section. 

\begin{acknowledgements}
This work was supported by National Key R\&D Program of China, Nos. 2017YFA0402400, 2022YFA1603200 and National Natural Science Foundation of China under grants U1931140 and 42074214, as well as Key Programs of the Chinese Academy of Sciences (QYZDJ-SSW-SLH050, QYZDJ-SSW-JSC028). We are grateful to Xiaoshu Wu from Sun Yat-sen University for providing MAVEN data. We thank LetPub (www.letpub.com) for its linguistic assistance during the preparation of this manuscript.
\end{acknowledgements}

\end{document}